\PassOptionsToPackage{unicode}{hyperref}
\PassOptionsToPackage{hyphens}{url}
\documentclass[
  12pt,
]{article}
\usepackage{xcolor}
\usepackage[top=2cm, bottom=2cm, left=2.5cm, right=2.5cm]{geometry}
\usepackage{amsmath,amssymb}
\usepackage{iftex}
\ifPDFTeX
  \usepackage[T1]{fontenc}
  \usepackage[utf8]{inputenc}
  \usepackage{textcomp} 
\else 
  \usepackage{unicode-math} 
  \defaultfontfeatures{Scale=MatchLowercase}
  \defaultfontfeatures[\rmfamily]{Ligatures=TeX,Scale=1}
\fi
\usepackage{lmodern}
\ifPDFTeX\else
\fi
\IfFileExists{upquote.sty}{\usepackage{upquote}}{}
\IfFileExists{microtype.sty}{
  \usepackage[]{microtype}
  \UseMicrotypeSet[protrusion]{basicmath} 
}{}
\makeatletter
\@ifundefined{KOMAClassName}{
  \IfFileExists{parskip.sty}{%
    \usepackage{parskip}
  }{
    \setlength{\parindent}{0pt}
    \setlength{\parskip}{6pt plus 2pt minus 1pt}}
}{
  \KOMAoptions{parskip=half}}
\makeatother
\usepackage{longtable,booktabs,array}
\usepackage{calc} 
\usepackage{etoolbox}
\makeatletter
\patchcmd\longtable{\par}{\if@noskipsec\mbox{}\fi\par}{}{}
\makeatother
\IfFileExists{footnotehyper.sty}{\usepackage{footnotehyper}}{\usepackage{footnote}}
\makesavenoteenv{longtable}
\usepackage{graphicx}
\makeatletter
\newsavebox\pandoc@box
\newcommand*\pandocbounded[1]{
  \sbox\pandoc@box{#1}%
  \Gscale@div\@tempa{\textheight}{\dimexpr\ht\pandoc@box+\dp\pandoc@box\relax}%
  \Gscale@div\@tempb{\linewidth}{\wd\pandoc@box}%
  \ifdim\@tempb\p@<\@tempa\p@\let\@tempa\@tempb\fi
  \ifdim\@tempa\p@<\p@\scalebox{\@tempa}{\usebox\pandoc@box}%
  \else\usebox{\pandoc@box}%
  \fi%
}
\def\fps@figure{htbp}
\makeatother
\NewDocumentCommand\citeproctext{}{}
\NewDocumentCommand\citeproc{mm}{%
  \begingroup\def\citeproctext{#2}\cite{#1}\endgroup}
\makeatletter
 \let\@cite@ofmt\@firstofone
 \def\@biblabel#1{}
 \def\@cite#1#2{{#1\if@tempswa , #2\fi}}
\makeatother
\newlength{\cslhangindent}
\newlength{\csllabelwidth}
\newenvironment{CSLReferences}[2] 
 {\begin{list}{}{%
  \setlength{\itemindent}{0pt}
  \setlength{\leftmargin}{0pt}
  \setlength{\parsep}{0pt}
  \ifodd #1
   \setlength{\leftmargin}{\cslhangindent}
   \setlength{\itemindent}{-1\cslhangindent}
  \fi
  \setlength{\itemsep}{#2\baselineskip}}}
 {\end{list}}
\usepackage{calc}

\newcommand{\CSLLeftMargin}[1]{\parbox[t]{\csllabelwidth}{\strut#1\strut}}
\newcommand{\CSLRightInline}[1]{\parbox[t]{\linewidth - \csllabelwidth}{\strut#1\strut}}

\usepackage{tabularx}
\usepackage{booktabs}
\usepackage{caption}
\usepackage{subcaption}
\usepackage{float}
\usepackage{graphicx}

\usepackage{array}
\usepackage{longtable}
\usepackage{fontspec}
\usepackage{multirow}
\usepackage{multicol}
\usepackage{colortbl}
\usepackage{hhline}
\newlength\Oldarrayrulewidth
\newlength\Oldtabcolsep
\usepackage{longtable}
\usepackage{array}
\usepackage{hyperref}
\usepackage{float}
\usepackage{wrapfig}
\usepackage{bookmark}
\IfFileExists{xurl.sty}{\usepackage{xurl}}{} 
\hypersetup{
  pdftitle={Instability in Patient Clustering: A Multiverse Analysis of Unsupervised Clustering in the CENTER-TBI cohort},
  pdfauthor={Sean R.E.A. Bagcik\^{}\{1,a\}; Aneeta Merlin Chacko\^{}\{2,a\}; Ewout W. Steyerberg\^{}\{2\}; Maarten van Smeden\^{}\{2\}; Andrew I. R. Maas\^{}\{3,4\}; Erik van Zwet\^{}\{1,b\}; Nicole S. Erler\^{}\{2,b\}; the SOPRANI collaborators \^{}\{+\}; and the CENTER-TBI participants and investigators \^{}\{+\}},
  hidelinks,
  pdfcreator={LaTeX via pandoc}}

\title{Instability in Patient Clustering: A Multiverse Analysis of Unsupervised Clustering in the CENTER-TBI cohort}
\author{Sean R.E.A. Bagcik\(^{1,a}\) \and Aneeta Merlin Chacko\(^{2,a}\) \and Ewout W. Steyerberg\(^{2}\) \and Maarten van Smeden\(^{2}\) \and Andrew I. R. Maas\(^{3,4}\) \and Erik van Zwet\(^{1,b}\) \and Nicole S. Erler\(^{2,b}\) \and the SOPRANI collaborators \(^{+}\) \and and the CENTER-TBI participants and investigators \(^{+}\)}
\date{}

\begin{document}
\maketitle

\(^{a,b}\) These authors contributed equally
\textsuperscript{+} Full list of author information is available at the end

\textbf{Affliations:}\\
\(^{1}\) Department of Biomedical Data Sciences, Leiden University Medical Center, Leiden, the Netherlands\\
\(^{2}\) Department of Data Science and Biostatistics, Julius Center for Health Sciences and Primary Care, UMC Utrecht, the Netherlands\\
\(^{3}\) Department of Neurosurgery, Antwerp University Hospital, Edegem, Belgium\\
\(^{4}\) Department of Translational Neuroscience, University of Antwerp, Belgium

\textbf{Corresponding Author}:
Aneeta Merlin Chacko, \href{mailto:a.m.chacko@umcutrecht.nl}{\nolinkurl{a.m.chacko@umcutrecht.nl}}

\nocite{*}

\section*{Abstract}\label{abstract}
\addcontentsline{toc}{section}{Abstract}

Understanding patient heterogeneity is key to improving prognostic modeling in traumatic brain injury (TBI). Unsupervised clustering is widely used to explore patterns in patient characteristics that may define subgroups. However, it involves a multitude of decisions, including the choice of algorithm, the distance metric, and the method used to determine the ``optimal'' number of clusters. The aim of this study is to investigate how these choices influence the resulting clustering solution.

We analyzed data from 4,509 patients enrolled in the Collaborative European NeuroTrauma Effectiveness Research in TBI (CENTER-TBI) study. K-medoids, agglomerative, and spectral clustering were applied in a complete \(3 \times 2 \times 2\) factorial design, in combination with Euclidean or Gower's distances, and silhouette score or gap statistic to choose the number of clusters. We investigated the agreement of clustering solutions with UpSet Plots and stability with the (adjusted) Rand index. Comparisons were made both across approaches using the original dataset and within approaches using bootstrap resampling.

Clustering results varied substantially depending on the analysis choices. The number of suggested clusters varied widely, from one to twenty-five. Adjusted Rand indices confirmed low concordance between methods. Moreover, none of the clustering solutions demonstrated discriminatory performance comparable to a supervised logistic regression model in classifying patient recovery illustrating the limited usefulness of clustering for this purpose. The high instability in clustering results compromises interpretability and underscores that such solutions should not be blindly interpreted as underlying structure.

\section{Introduction}\label{introduction}

Patients presenting with traumatic brain injury (TBI) are generally classified according to the Glasgow Coma Scale (GCS), which is widely used to assess the level of consciousness in patients. Its sum score ranges from 3 (completely unresponsive) to 15 (completely responsive) which is often categorized as mild (13-15), moderate (9-12) or severe (3-8) levels of clinical severity of TBI. While low GCS scores indicate severe injury, patients with similar GCS scores can have widely divergent functional outcomes. This makes the GCS insufficient as a predictive classifier on its own {[}\citeproc{ref-manley2025new}{1}{]}. Additionally, it does not capture the specific features of pathophysiology in individual patients. The multidimensionality of TBI requires a more comprehensive and detailed classification than just the GCS {[}\citeproc{ref-manley2025new}{1}{]}.

Unsupervised clustering is sometimes applied to identify distinct disease phenotypes and prognostic profiles for clinical outcomes in diverse patient populations, such as those with TBI {[}\citeproc{ref-tas2024unsupervised}{2}{]}. Unsupervised clustering methods group patients based on patterns in their clinical, demographic, or imaging features without relying on outcome labels. Patients within a cluster are more ``similar'' to one another than to patients in other clusters. However, this broad aim leaves much room for implementation, specifically regarding how similarity (i.e., the ``distance'' between patients and between clusters) should be defined, how intra- and inter-cluster distances should be weighed against each other and when clusters should be split or merged. Decisions involved in clustering analyses range from data cleaning and pre-processing, feature selection and algorithm selection to hyperparameter tuning, and reflect the broader issue of ``researcher degree of freedom'': the many plausible analytical pathways that are available when performing clustering. Borges' ``garden of forking paths'' {[}\citeproc{ref-gelman2013forkingpaths}{3}{]}, captures this challenge: reasonable but different analytic decisions can lead to vastly different results and conclusions. In a clinical context the usefulness of clustering for risk stratification is doubtful, since unsupervised clustering methods do not incorporate relationships between features or clusters and the clinical outcome.

A previous unsupervised clustering analysis included 4,509 TBI patients from the CENTER-TBI study {[}\citeproc{ref-gravesteijn2020toward}{4}{]}. Four distinct patient clusters were identified with the K-medoids algorithm {[}\citeproc{ref-pam1990}{5}{]}, using the silhouette score {[}\citeproc{ref-rousseeuw1987silhouettes}{6}{]} to determine the optimal number of clusters and a version of Gower's distance {[}\citeproc{ref-gower1971similarity}{7}{]}. While this selection of these four clustering techniques is likely a sensible approach, many other choices could be equally sensible. In this study, we explore the extent to which alternative choices in unsupervised clustering impact the results. We aim to reveal the uncertainty introduced by the researcher degrees of freedom. We apply a multiverse-style approach in which we follow Gravesteijn et al. {[}\citeproc{ref-gravesteijn2020toward}{4}{]} in their choice of data pre-processing and feature selection, but systematically explore several options for the choice of the clustering, distance metric and method to determine the optimal number of clusters.

We compare three common unsupervised clustering algorithms (k-medoids, agglomerative clustering, and spectral clustering), under two alternative distance metrics (Euclidean distance and Gower's distance) and use two approaches to determine the optimal number of clusters (silhouette values and the gap statistic). These methods were chosen to cover a range of widely used clustering strategies. To investigate the stability of the resulting clustering solutions, we compare the number, size, and means of clusters and evaluate the (adjusted) Rand index, both between clustering strategies as well as within clustering strategies using resampling approaches.

The remainder of this article is organized as follows: In Section 2, we introduce the CENTER-TBI data and the original clinical research question, with the data pre-processing. The clustering algorithms and choices that form part of our multiverse analysis are presented in Section 3, followed by a description of the methods used to evaluate and compare the clustering solutions, in Section 4. The results of the multiverse clustering analysis of the CENTER-TBI data are presented in Section 5 and Section 6 concludes with a discussion.

\section{The CENTER-TBI data}\label{the-center-tbi-data}

In this study, we work with the data from the Collaborative European NeuroTrauma Effectiveness Research in Traumatic Brain Injury (CENTER-TBI) study, a large, multicenter, prospective observational cohort of patients with traumatic brain injury (TBI) {[}\citeproc{ref-maas2015center}{8},\citeproc{ref-steyerberg2019casemix}{9}{]}. The CENTER-TBI study is a landmark European research initiative aimed at closing major gaps in TBI research, including poor classification systems, lack of predictive tools, and limited evidence for treatment strategies. Between December 2014 and December 2017, approximately 4500 patients were included in 63 centers across 18 countries {[}\citeproc{ref-steyerberg2019heterogeneity}{10}{]}. Inclusion criteria of the CENTER-TBI study were a clinical diagnosis of TBI, presentation within 24 hours after injury and an indication for computed tomography. The CENTER-TBI cohort includes patients across the full spectrum of traumatic brain injury severity, from mild to severe, and encompasses a wide range of ages, comorbidities, and clinical presentations. Clustering is a tempting approach used for such heterogeneous datasets because it allows for exploration of underlying structures and groups of individuals without relying on pre-defined labels.

In line with Gravesteijn et al. {[}\citeproc{ref-gravesteijn2020toward}{4}{]}, we used data from all 4509 available patients and included the same seven baseline characteristics and twelve imaging features (Table \ref{tab:tab1}). Since its release, the CENTER-TBI dataset has been updated to ensure high data quality, inclusion of late-available variables, and correction of inconsistencies. In the current analysis, we work with CENTER TBI version 3.1 (released on 31 January 2025), which has minor differences compared to the version used by Gravesteijn et al. {[}\citeproc{ref-gravesteijn2020toward}{4}{]} (version 1.0). The commonly used clinical outcome is the extended Glasgow Outcome Scale (GOSE) measured at 6 months after injury. GOSE is an ordinal scale with eight categories, ranging from death (1) to upper good recovery (8). Following standard practice in traumatic brain injury research, we dichotomised GOSE into unfavourable (GOSE \(< 5\)) and favourable (GOSE \(\ge 5\)) outcomes. Approximately 25\% of patients in the dataset experienced an unfavourable outcome.

\subsection{Data pre-processing}\label{data-pre-processing}

The data pre-processing followed the steps described by Gravesteijn et al. {[}\citeproc{ref-gravesteijn2020toward}{4}{]}. Missing values were imputed via single imputation by chained equations using the \textbf{mice}-package version 3.17.0 {[}\citeproc{ref-vanbuuren2018flexible}{11}{]} in R {[}\citeproc{ref-R-base}{12}{]} version 4.4.1. All variables shown in Table \ref{tab:tab1}, i.e., the 19 variables included in the clustering, as well as the clinical outcome (GOS-E at 6 months), were used in the imputation. We used predictive mean matching with distance aided donor selection (midastouch) for continuous variables and logistic regression for categorical variables. We deemed single imputation sufficient for our purpose as we only required a representative dataset to compare the clustering methods. Principal Component Analysis was used to reduce the twelve imaging features to four principal components, following the approach of Gravesteijn et al. {[}\citeproc{ref-gravesteijn2020toward}{4}{]}. Numerical features were scaled to have mean zero and standard deviation equal to one to standardize their contribution to distance calculations, while categorical features with \(n\) levels were encoded as integer values ranging from \(0\) to \(n-1\). Although this is a common approach when incorporating categorical features in clustering, it is important to note that depending on the distance metric used, variables with many categories may exert more influence on the clustering than those with only a few categories. Also, note that this may be reasonable for ordinal features, but it makes no sense for nominal features.

\section{Clustering Framework}\label{clustering-framework}

We implemented a multiverse-style analysis and varied key components of the clustering pipeline: the algorithm, the distance metric, and the method for selecting the number of clusters. In this section we describe the specific choices made in each of these domains.

\subsection{Clustering Strategies}\label{clustering-strategies}

Clustering algorithms can be broadly categorized into three families based on how they define and construct clusters: centroid-based, hierarchical, and graph-based methods. Each family makes different assumptions about cluster shape, density, and separation. In this study, we selected one representative algorithm from each family: k-medoids clustering (centroid-based), agglomerative clustering (hierarchical), and spectral clustering (graph-based). These algorithms were chosen for their conceptual diversity, widespread use in applied research, and compatibility with precomputed distance metrics. Additionally, all three require the number of clusters to be specified in advance, making them suitable for systematic comparison under controlled conditions.

\subsubsection{Centroid based clustering}\label{centroid-based-clustering}

K-medoids clustering is a centroid-based partitioning method that minimizes the total dissimilarity between observations and their assigned cluster medoid --- an actual data point that serves as the cluster center. It is considered to be well-suited for clinical data due to its robustness to outliers and compatibility with arbitrary dissimilarity measures {[}\citeproc{ref-hastie2009elements}{13}{]}. Formally, the objective is to minimize the total within-cluster dissimilarity:
\begin{equation}
\label{eq:kmed}
\sum_{k=1}^K \sum_{x_i \in C_k} {d(x_i, m_k)},
\end{equation}
where \(K\) is the pre-specified number of clusters, \(C_k\) denotes the \(k^{th}\) cluster, \(x_i\) is the vector of features for the \(i^{th}\) patient, \(m_k \in C_k\) is the medoid of that cluster, and \(d(\cdot,\cdot)\) is a precomputed dissimiliarity metric (e.g., Euclidean or Gower's distance). We used the alternating optimization strategy {[}\citeproc{ref-hastie2009elements}{13}{]} as implemented in the \emph{scikit-learn-extra} Python package version 0.3.0 {[}\citeproc{ref-scikit-learn}{14}{]} to solve this optimization problem. It iteratively assigns points to the nearest medoid and updates medoids by selecting the point within each cluster that minimizes the total dissimilarity. The Partitioning Around Medoids (PAM) algorithm is computationally more intensive, making it unfeasible to use in our application.

\subsubsection{Hierarchical clustering}\label{hierarchical-clustering}

Agglomerative clustering is a bottom-up hierarchical clustering method that builds a nested tree (dendrogram) of clusters by iteratively merging the two closest clusters based on a specified linkage criterion {[}\citeproc{ref-jain1999data}{15}{]}. Starting from singleton clusters, at each step, the pair of clusters \((C_{k}, C_{l})\) with the smallest inter-cluster dissimilarity is merged, where the dissimilarity is defined by a linkage function \(D(C_{k}, C_{l})\). In our analysis, we used average linkage, which defines the dissimilarity between two clusters \(C_{k}\) and \(C_{l}\) as the average pairwise dissimilarity between their elements:
\begin{equation}
\label{eq:agg}
D(C_{k}, C_{l}) = \frac{1}{n_{k} n_{l}} \sum_{i\in C_{k}} \sum_{j \in C_{l}} d(x_i, x_j),
\end{equation}
where \(d(x_i, x_j)\) is the dissimilarity between observations \(x_i\) and \(x_j\) (e.g., Euclidean or Gower's distance) and \(n_{k}\) and \(n_{l}\) are the number of elements in clusters \(C_{k}\) and \(C_{l}\), respectively. Average linkage represents a compromise between the extremes of single and complete linkage, producing clusters that are moderately compact while maintaining reasonable separation {[}\citeproc{ref-hastie2009elements}{13}{]}. Contrary to Ward linkage, the default in \emph{scikit-learn} version 1.4.2 {[}\citeproc{ref-scikit-learn}{14}{]}, the average linkage is compatible with arbitrary dissimilarity matrices, making it more suitable for our multiverse framework.

\subsubsection{Graph-based clustering}\label{graph-based-clustering}

Spectral clustering is a graph-based method. A similarity graph \(\textbf{S}=\{s_{ij}\}\) is constructed in which patients are the nodes. Two nodes \(x_i\) and \(x_j\) are connected by an edge if their pairwise similarity \(s_{ij}\ge 0\), and these edges are additionally weighted by \(s_{ij}\). The pairwise similarities are obtained from distance metrics. For the Euclidean distance, we constructed the similarity matrix \(\textbf{S}\) using a radial basis function kernel:
\begin{equation}
\label{eq:spec2}
s_{ij}=\text{exp} \Biggl\{ -\frac{d_E(x_i,x_j)^2}{2\sigma^2}\Biggl\},
\end{equation}
where \(d_E(x_i,x_j)\) is the Euclidean distance (equation \ref{eq:eucl}) and \(\sigma\) is a scaling parameter set to the median pairwise distance. Since Gower's distance is restricted to values between \(0\) and \(1\), it can be converted to a similarity measure as \(s_{ij}=1-d_G(x_i,x_j)\), where \(d_G(x_i,x_j)\) is the Gower's distance (equation \ref{eq:gow}).

The goal is then to divide the similarity graph into clusters such that points within the same cluster are strongly connected (high similarity) and points in different clusters are weakly connected (low similarity) {[}\citeproc{ref-hastie2009elements}{13}{]}. This can be achieved by clustering (e.g.~with k-means) the eigenvectors that correspond to the \(K\) smallest non-zero eigenvalues of the normalized graph Laplacian matrix,
\begin{equation}
\label{eq:spec1}
\textbf{L}=\textbf{I}-\textbf{D}^{-1/2} \textbf{SD}^{-1/2},
\end{equation}
where \(\textbf{D}\) is the diagonal degree matrix with \(\textbf{D}_{ii}=\sum_j s_{ij}\) that describes the connectivity of each observation \(x_i\) to all others, and \(\textbf{I}\) is the identity matrix. Normalization helps to prevent nodes with few edges with large weights dominate the clustering.

This approach effectively embeds the data into a lower-dimensional space where the cluster structure is more pronounced. While k-medoids and agglomerative clustering operate directly in the original feature space, spectral clustering can capture more complex, non-convex structures, but it is sensitive to how the similarity graph is constructed. Xu and Tian {[}\citeproc{ref-xu2015comprehensive}{16}{]} provide a comprehensive survey highlighting the advantages and disadvantages of k-medoids, agglomerative and spectral clustering.

We implemented spectral clustering using the \emph{scikit-learn} library in Python, which supports precomputed affinity matrices {[}\citeproc{ref-scikit-learn}{14}{]}. The number of eigenvectors was set to the default, which equals the specified number of clusters.

\subsection{Distance Measures}\label{distance-measures}

Distance measures are a fundamental component of clustering algorithms because they determine how similarity between observations is quantified. Choosing an appropriate distance measure is not straightforward when features differ in scale (e.g., age in years vs GCS score) or differ in measurement levels (e.g., continuous vs categorical). Many distance measures have been proposed {[}\citeproc{ref-deza2009encyclopedia}{17}{]}. We focus on two of the most commonly used measures: the Euclidean distance and Gower's distance.

The Euclidean distance is one of the most widely used measures of dissimilarity for continuous variables {[}\citeproc{ref-hastie2009elements}{13}{]}. It represents the straight-line distance between two points in a multidimensional space and is defined as
\begin{equation}
\label{eq:eucl}
d_E(x_i,x_j)=\sqrt{\sum_{p=1}^P (x_{ip}-x_{jp})^2},
\end{equation}
where \(x_{ip}\) and \(x_{jp}\) denote the value of the \(p^{th}\) feature of patients \(i\) and \(j\) respectively. Since the Euclidean distance is sensitive to differences in scale, meaning that variables with larger ranges can dominate the distance calculation, all continuous variables were standardized during the pre-processing. Naturally, the Euclidean distance treats a one-unit difference in any feature as equally important, regardless of the differences in the features' scales or the clinical relevance of a one-unit difference. After scaling, a difference of one standard deviation is considered equally important, which may improve comparability across continuous features. However, even after scaling, the calculated distances should be interpreted as relative measures of similarity rather than exact representations of actual distances between patients as scaling cannot fully resolve the fundamental issue of comparing variables measured on different scales.

While the Euclidean distance is appropriate for numeric features, it is not well-suited for categorical variables. Encoding categories as numeric values (e.g., 0,1,2) introduces an arbitrary order for nominal variables and assumes equal spacing between categories. Since there is no true notion of distance between patients, this can lead to misleading results. Nevertheless, this practice remains common in applied work, and we therefore include it in our multiverse analysis.

Gower's distance is a dissimilarity measure designed for datasets with mixed variable types, such as continuous, ordinal and nominal features. For a pair of patients \(x_i\) and \(x_j\), Gower's distance is computed as the average of variable-specific dissimilarities {[}\citeproc{ref-gower1971similarity}{7}{]}:
\begin{equation}
\label{eq:gow}
d_G(x_i,x_j)=\frac{1}{P}\sum_{p=1}^P d_{ijp},
\end{equation}
where \(P\) is the number of features, \(R_p\) is the range of feature \(x_p\), and \(d_{ijp}=\frac{|x_{ip}-x_{jp}|}{R_p}\) for continuous \(x_p\) and
\(d_{ijp} =
\begin{cases}
0 & \text{if } x_{ip} = x_{jp} \\
1 & \text{otherwise}
\end{cases}\).

This normalization ensures that all variables contribute equally regardless of their scale or type. Gower's distance treats categorical variables equally regardless of the number of categories, since it only evaluates whether values are the same or different rather than measuring the magnitude of difference. This makes Gower's distance an attractive choice for clinical data, where features often include a mix of continuous and categorical variables.

\subsection{Choosing the number of clusters}\label{choosing-the-number-of-clusters}

Many clustering techniques require the user to pre-specify the number of clusters. As this number can often not be treated as a known property, data-driven methods have been developed to determine the optimal number of clusters. Since the criterion that is optimized differs between methods, the ``optimal'' number of clusters depends on the chosen method. We considered two popular approaches: the silhouette index {[}\citeproc{ref-rousseeuw1987silhouettes}{6}{]} and the gap statistic {[}\citeproc{ref-tibshirani2001gap}{18}{]}.

The silhouette index quantifies the quality of a clustering solution, based on how similar observations are to the other observations in their own cluster (cohesion) compared to other clusters (separation). Specifically, the silhouette value of subject \(i\) is calculated as:
\begin{equation}
\label{eq:sil}
s(i)=\frac{b(i)-a(i)}{\text{max}\{a(i),b(i)\}},
\end{equation}
where \(a(i)=\frac{1}{|C_i|-1}\sum_{j \in C_i,i \ne j }d(i,j)\) is the average dissimilarity of \(x_i\) to all other points in its own cluster and \(b(i)=\text{min}_{C_j\ne C_i}\frac{1}{|C_j|}\sum_{j\in C_j}d(i,j)\) is the lowest average dissimilarity of \(x_i\) to points in other clusters. In our analysis, we calculated the silhouette index using the same dissimilarity measure as used in the clustering algorithm, i.e., either the Euclidean or Gower's distance. If subject \(i\) is the only member of its cluster, then its silhouette index is set to zero. The silhouette index ranges from −1 to 1, with 1 indicating perfect within-cluster cohesion and −1 indicating that observations are likely assigned to a wrong cluster. The ``global silhouette index'' is then defined as the average of all the individual silhouette values {[}\citeproc{ref-rousseeuw1987silhouettes}{6}{]} and is calculated as
\(S=\frac{1}{N}\sum_{i=1}^N s(i)\). To find the optimal number of clusters, the silhouette index is calculated for clustering solutions with a range of candidate values for \(K\) and the optimal number of clusters is the one that maximizes the silhouette index. The silhouette index requires at least two clusters to compute meaningful within and between cluster dissimilarities. Consequently, evaluations using this metric begin at a minimum of two clusters making it impossible to find that there are no clusters present (i.e., K=1) when using the silhouette index. We implemented the silhouette index in Python using the \textbf{silhouette\_score} function from the \emph{sklearn.metrics} module {[}\citeproc{ref-scikit-learn}{14}{]}.

The gap statistic is a resampling-based method to determine the number of clusters that compares the within-cluster dispersion of a clustering solution to the expected dispersion under an appropriate reference null distribution {[}\citeproc{ref-tibshirani2001gap}{18}{]}. There are various options for the reference distribution, with the simplest being the uniform distribution for numerical features and discrete uniform distribution for categorical features. To calculate the gap statistic for a given number of clusters K, first the within-cluster dispersion is computed as
\begin{equation}
\label{eq:gap1}
W_k=\sum_{i=1}^K\frac{1}{2|C_k|}\sum_{x_i,x_j \in C_k}d(x_i,x_j),
\end{equation}
where \(C_k\) denotes the set of observations in cluster \(k\) and \(d(\cdot,\cdot)\) is the chosen distance measure. Naturally, \(W_K\) decreases with increasing \(K\), but beyond a certain point the decrease flattens. Thus, to determine the optimal number of clusters, Tibshirani et al. {[}\citeproc{ref-tibshirani2001gap}{18}{]} propose to compare the observed value of \(\text{log}(W_K)\) to its expected value under the reference distribution. The gap statistic is then defined as
\begin{equation}
\label{eq:gap2}
\text{Gap}_n(K)=\mathbb{E}^*_n \{\text{log} (W^*_K)\} - \text{log} ( {W_K} ), 
\end{equation}
where \(\mathbb{E}^*_n\{\text{log} (W^*_K)\}\) is the expected log-dispersion based on \(n\) samples (simulations) from the reference distribution. A larger gap value indicates stronger evidence for clustering beyond a random structure. The optimal number of clusters is the smallest value of \(K\) for which the gap statistic lies within one-standard error of the gap at \(K+1\) as proposed by Tibshirani et al. {[}\citeproc{ref-tibshirani2001gap}{18}{]}. In contrast to the silhouette index, the gap statistic can be evaluated for a single-cluster solution and can therefore determine a one cluster solution (i.e., no clustering) to be the best solution for the data (under the chosen clustering algorithm and distance metric).

The gap statistic was implemented in Python through a custom function that compares within-cluster dispersion on the observed data to that of reference datasets {[}\citeproc{ref-tibshirani2001gap}{18}{]}. 200 reference datasets were generated by simulating continuous variables from a uniform distribution across their observed ranges and categorical variables from a discrete uniform distribution over the observed categories. The gap statistic and its standard error were estimated using these reference datasets.

\section{Evaluation of clustering solutions}\label{evaluation-of-clustering-solutions}

The validation of solutions from unsupervised clustering is not straightforward because ``true labels'' do not exist {[}\citeproc{ref-von2012clustering}{19}{]}. In the absence of a ground truth, directly assessing the quality of the clustering solutions may not be possible, thus evaluation relies on indirect approaches such as stability analyses.

\subsection{Clustering Stability}\label{clustering-stability}

Cluster evaluation through stability refers to assessing the consistency of clustering results when applied to perturbed versions of the original dataset or to independent samples from multiple studies. If the clustering is stable, the clusters from the original data will be preserved in the different versions of data. A common approach is resampling-based stability analysis using techniques such as bootstrapping and comparing the clustering results across re-sampled datasets to quantify the degree of agreement {[}\citeproc{ref-liu2022stability}{20}{]}.

We used the adjusted Rand index (ARI) to quantify the clustering stability because it is commonly employed and straightforward to interpret. For two clustering solutions X and Y, the pair-confusion matrix classifies each patient pair into one of four categories:\\
- \emph{a}: both X and Y assign the pair to the same cluster,\\
- \emph{b}: only X assigns the pair to the same cluster,\\
- \emph{c}: only Y assigns the pair to the same cluster, and\\
- \emph{d}: both X and Y assign the pair to different clusters,

where \emph{a} and \emph{d} indicate agreement, and \emph{b} and \emph{c} imply disagreement. Let \(a,b,c \text{ and } d\) denote the number of patient pairs in each of these categories. The Rand index (RI) is defined as \(RI     = \frac{(a+d)}{a+b+c+d}\). It is the probability that the two clustering solutions agree on a randomly selected pair of patients. The Rand index is large when both solutions have few, large clusters since that makes chance agreement more likely. To adjust the Rand index for this, it can be compared to the expected agreement that would arise from randomly assigning patients to clusters. This random assignment follows the hypergeometric distribution of the cells in the pair-confusion matrix {[}\citeproc{ref-hubert1985comparing}{21}{]}. The Adjusted Rand Index (ARI) is calculated as:
\begin{equation}
\label{eq:ari}
\text{ARI}=\frac{\text{RI}-\mathbb{E}(\text{RI})}{1-\mathbb{E}(\text{RI})},
\end{equation}

where \(\mathbb{E}(\text{RI})\) is the expected value of RI under random assignment. The ARI is bounded between \(-0.5\) and \(1\). If the unadjusted Rand index equals its expected value under randomness of cluster assignment, the ARI will be zero. In that case, there is no meaningful agreement beyond what would be expected by chance alone. An ARI of 1 indicates perfect agreement between the two clustering solutions. We use the ARI to compare clustering solutions resulting from different modelling choices as well as to assess the stability of a single clustering method when applied to bootstrap samples. To that end, we drew 200 bootstrap samples from the data and for each bootstrap sample we applied the clustering method to obtain cluster assignments. For all possible pairs of bootstrap samples, we computed the adjusted and unadjusted Rand indices. We report the averages of these indices across all bootstrap sample pairs for each clustering method.

\subsection{Clustering Agreement}\label{clustering-agreement}

Agreement between clustering approaches can be visualized using Upset plots {[}\citeproc{ref-lex2014upset}{22}{]}. For a dataset with N patients, there are \(\frac{N(N-1)}{2}\) unordered patient pairs. Each clustering method will either group a patient pair together or separate them. Thus, each clustering solution can be associated with the set of pairs it joins. The exclusive intersection of a subset of clustering methods includes all patient pairs that are grouped together by all methods in the subset and by none of the other methods. A large exclusive intersection suggests high agreement within that subset of methods. The Upset plot visualizes these exclusive intersections, their size, and the number of joined pairs for each clustering method.

Moreover, we use the Adjusted Rand Index (ARI) to quantify pairwise agreement between clustering solutions. Together, the UpSet plots and ARI values offer both a graphical and numerical assessment of the extent to which clustering results align across algorithms, distance metrics, and cluster number selection strategies.

\subsection{Predictive Value of the Identified Clusters}\label{predictive-value-of-the-identified-clusters}

Internal cluster validation metrics such as cluster stability may favor clusters that are compact and well-separated, but are not necessarily clinically meaningful. In clinical applications, such as the motivating TBI example, clusters might identify patient groups based on (yet unknown) covariate patterns, with the hope that these groups can be used for risk stratification. We investigated whether the clustering solutions can discriminate patients with different risk for a poor clinical outcome. Although this is not actually a method for assessing clustering validity, it is common practice in applied research, such as in case of TBI {[}\citeproc{ref-gravesteijn2020toward}{4}{]}. Note that the absence of an association should not be interpreted as evidence against the validity of the identified clusters, as clusters may capture structure not reflected in the tested outcomes.

Our clinical outcome of interest is the extended Glasgow Outcome Scale (GOSE) at 6 months after the injury. For each clustering solution, the expected probability of an unfavourable outcome for patients in a particular cluster was calculated as the observed proportion of unfavourable outcomes in that cluster. We quantified the discriminative ability of a clustering solution via the Area Under the Receiver Operating Characteristic Curve (AUC) and used bootstrap resampling of the original dataset (B = 200) to adjust for optimism and to obtain confidence intervals {[}\citeproc{ref-steyerberg2001internal}{23}{]}. The optimism was calculated as:
\begin{equation}
\label{eq:opt}
\text{Optimism} = \frac{\sum_{b=1}^{B}\text{AUC}_{b,\text{bootstrap}} - \text{AUC}_{b,\text{original}}}{B},
\end{equation}
where \(\text{AUC}_{b,\text{bootstrap}}\) refers to the AUC of the clustering solution determined in bootstrap sample b when applied to that same sample and \(\text{AUC}_{b,\text{original}}\) is the AUC of the same clustering solution evaluated on the original dataset. The optimism corrected AUC is then calculated as:
\begin{equation}
\label{eq:aucopt}
\text{AUC}_{\text{Optimism}-\text{Corrected}} = \text{AUC}_{\text{Apparent}} - \text{Optimism}, 
\end{equation}
where \(\text{AUC}_{\text{Apparent}}\) refers to the AUC of the clustering solution determined in the original data when applied to the original dataset.

Since risk stratification is classically a task for supervised methods, we additionally estimated a logistic regression model to obtain a reference value for the AUC. Previous research in a similar setup has shown that more complex machine learning models do not necessarily improve predictive performance beyond what logistic regression can achieve {[}\citeproc{ref-van2016modern}{24},\citeproc{ref-gravesteijn2020machine}{25}{]}. The logistic regression model included all features used in the clustering and assumed independent, linear associations with the log-odds. An optimism-corrected AUC was again obtained using bootstrap samples.

\section{Clustering of the CENTER-TBI data}\label{clustering-of-the-center-tbi-data}

We considered twelve clustering strategies, determined by combining the three clustering algorithms in combination with two distance metrics and two cluster selection methods in a full factorial design. For each strategy, we evaluated solutions for a range of cluster numbers (\(k = 2 \text{ to } 25\) for the silhouette score and \(k = 1 \text{ to } 25\) for the gap statistic) and selected the optimal number of clusters as described in Section 3.3.

\begin{figure}

{\centering \includegraphics[width=0.9\linewidth]{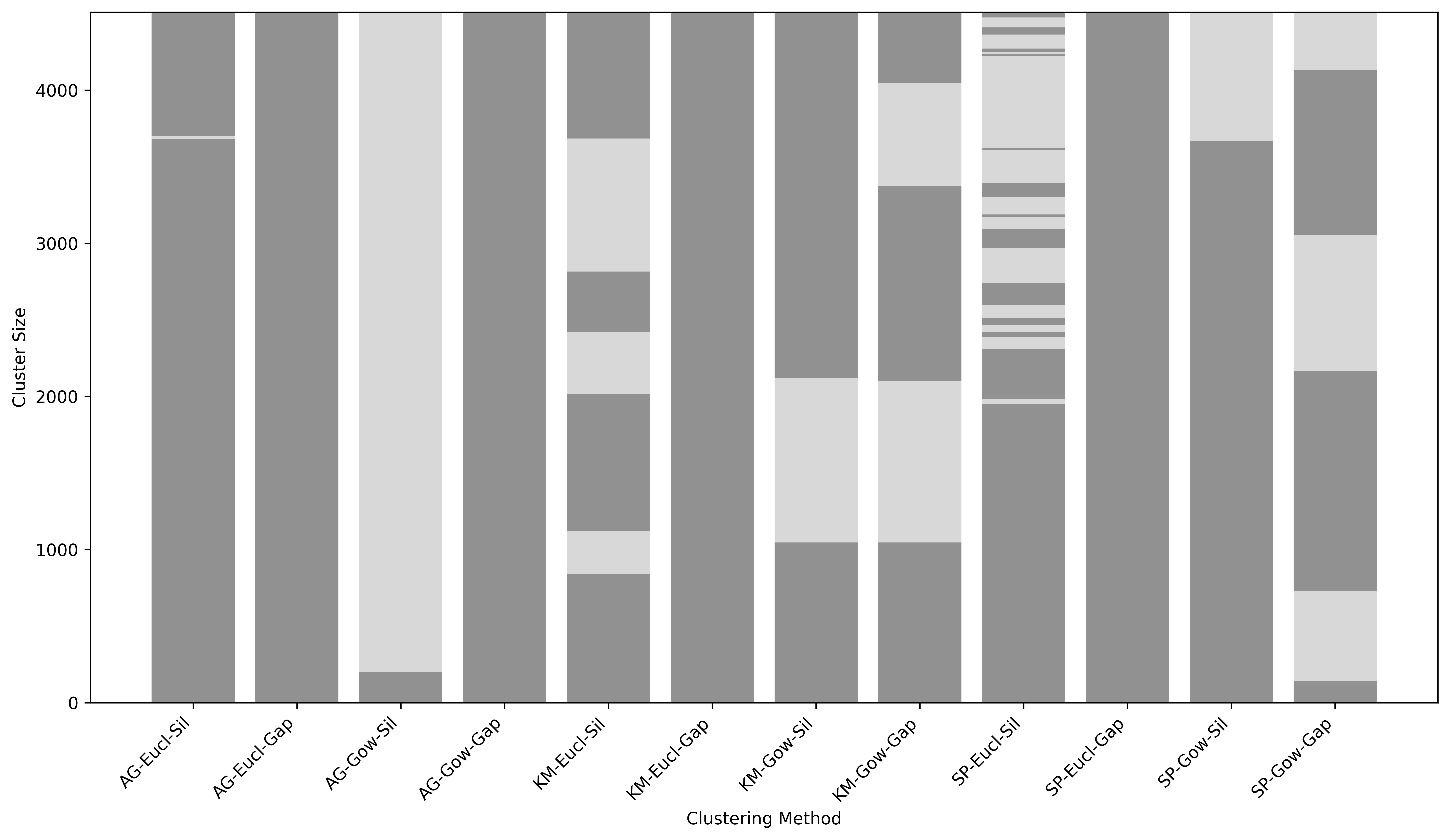} 

}

\caption{Cluster size (and number of clusters) across methods. Number of clusters between 1 and 25 were identified. \\\footnotesize{Note: The colours are purely illustrative and used for visual separation. Clusters sharing the same colour across different methods do not correspond to the same set of patients and should not be interpreted as corresponding groups. \\Abbreviations: AG = agglomerative clustering; KM = K-Medoids clustering; SP = spectral clustering; Eucl = Euclidean distance; Gow = Gower's distance; Sil = silhouette value; Gap = gap statistic}}\label{fig:cluster-size}
\end{figure}

Large differences were found in the solutions across clustering methods with the optimal number of clusters ranging from one to 25 (the maximum number considered) and the size of the largest cluster ranging from 869 to 4509 (the full sample size) (see Figure \ref{fig:cluster-size}). For four of the six algorithm-distance combinations, the gap statistic favoured a solution with a single cluster, indicating the absence of a clustering structure in the data. For spectral clustering with Gower's distance, the silhouette index (SP-Gow-Sil) found two clusters, where the smaller cluster contained only 839 patients, while the silhouette index (SP-Eucl-Sil) found 25 clusters, with the smallest cluster containing only 10 patients.

\subsection{Agreement between different clustering strategies}\label{agreement-between-different-clustering-strategies}

\begin{figure}[H]
\subfloat[Agreement between clustering solutions using Gower's distance\label{fig:fig2-1}]{\includegraphics[width=0.85\linewidth]{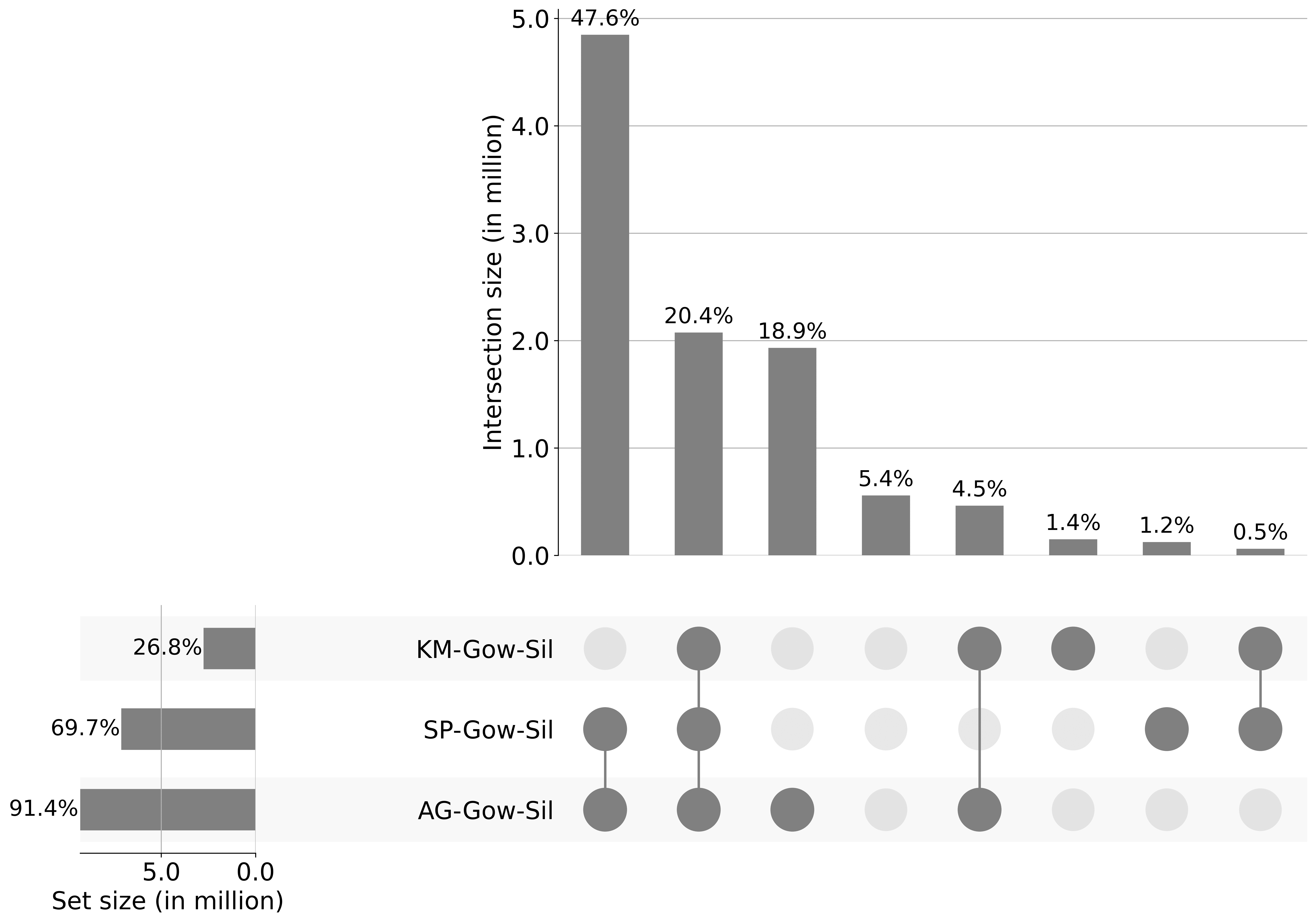} }\newline\subfloat[Agreement between clustering solutions using the Euclidean distance\label{fig:fig2-2}]{\includegraphics[width=0.85\linewidth]{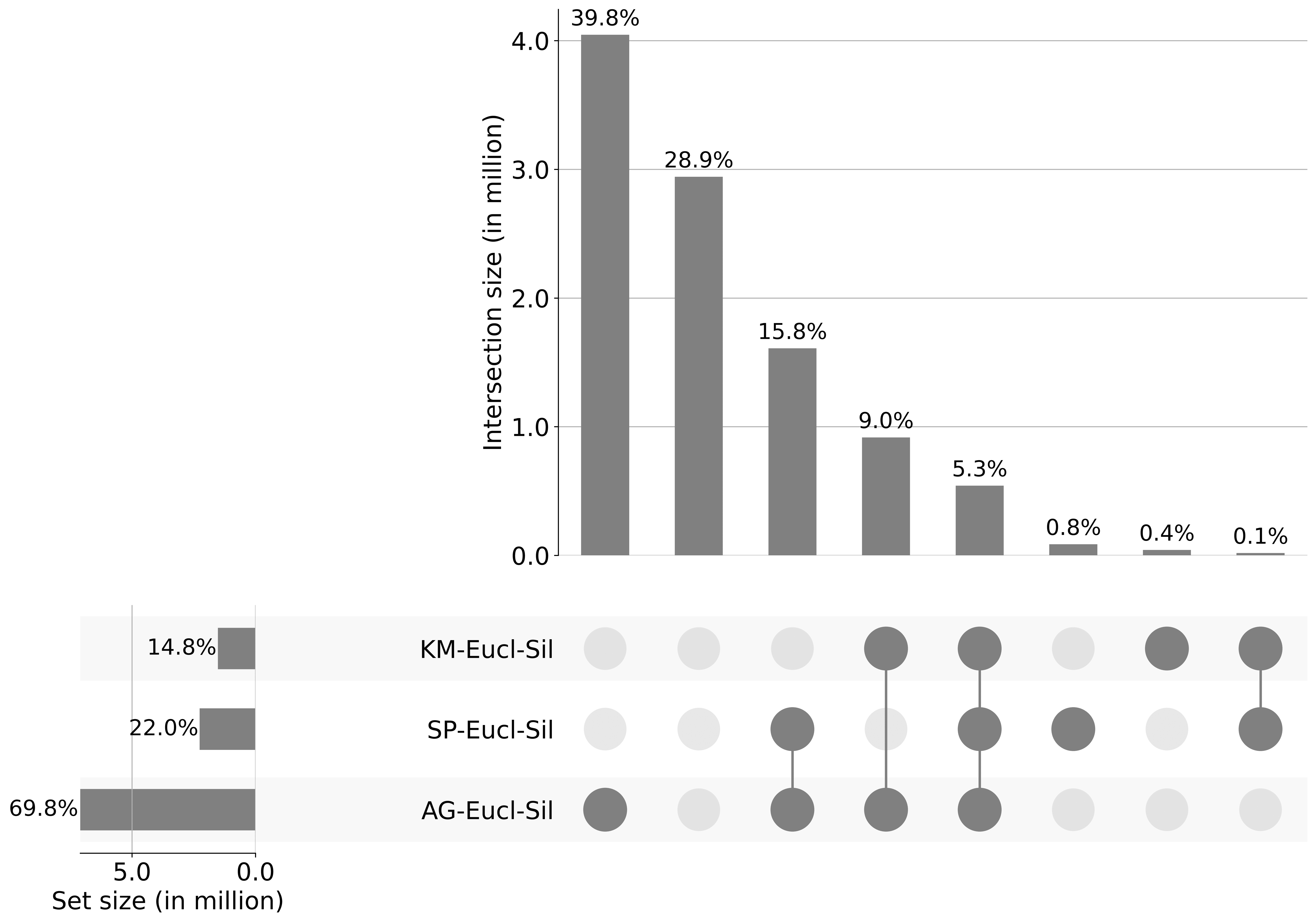} }\newline\caption{Agreement between clustering solutions when using the silhouette index. \\\footnotesize{Note: Each row represents a clustering strategy, and each vertical bar corresponds to the exclusive intersection indicated by the filled circles in that column. The percentage labels indicate the proportion of pairs of patients placed together in a cluster by each of the methods for which the circle is filled. For example, a column with all three methods selected represents patient pairs joined by all three clustering methods, while a column with only two filled circles shows agreement between those two methods only. Horizontal bars on the left show the total number of joined pairs for each individual method.}}\label{fig:fig2}
\end{figure}

We used Upset plots to visualize comparisons between the three clustering algorithms for a given combination of distance metric and method determining the number of clusters (Figure \ref{fig:fig2}). For Gower's distance, 47.6\% of all pairs of patients were grouped together by agglomerative and spectral clustering (Figure \ref{fig:fig2-1}), but the same methods grouped only 15.8\% of patient pairs together when using the Euclidean distance (Figure \ref{fig:fig2-2}). When using Gower's distance, the three clustering algorithms agreed on only 20.4\% of patient pairs being assigned to the same cluster, whereas with Euclidean distance, agreement dropped to just 5.3\% of patient pairs. The UpSet plots indicate stronger agreement between the clustering solutions for Gower's distance, which may be due to the number of clusters being lower for these strategies compared to when using the Euclidean distance. We did not present UpSet plots for the gap statistic--based solutions because this method frequently selected a single-cluster solution. In such cases, all patients are assigned to the same cluster, making agreement comparisons across solutions uninformative.

\begin{figure}

{\centering \includegraphics[width=0.95\linewidth]{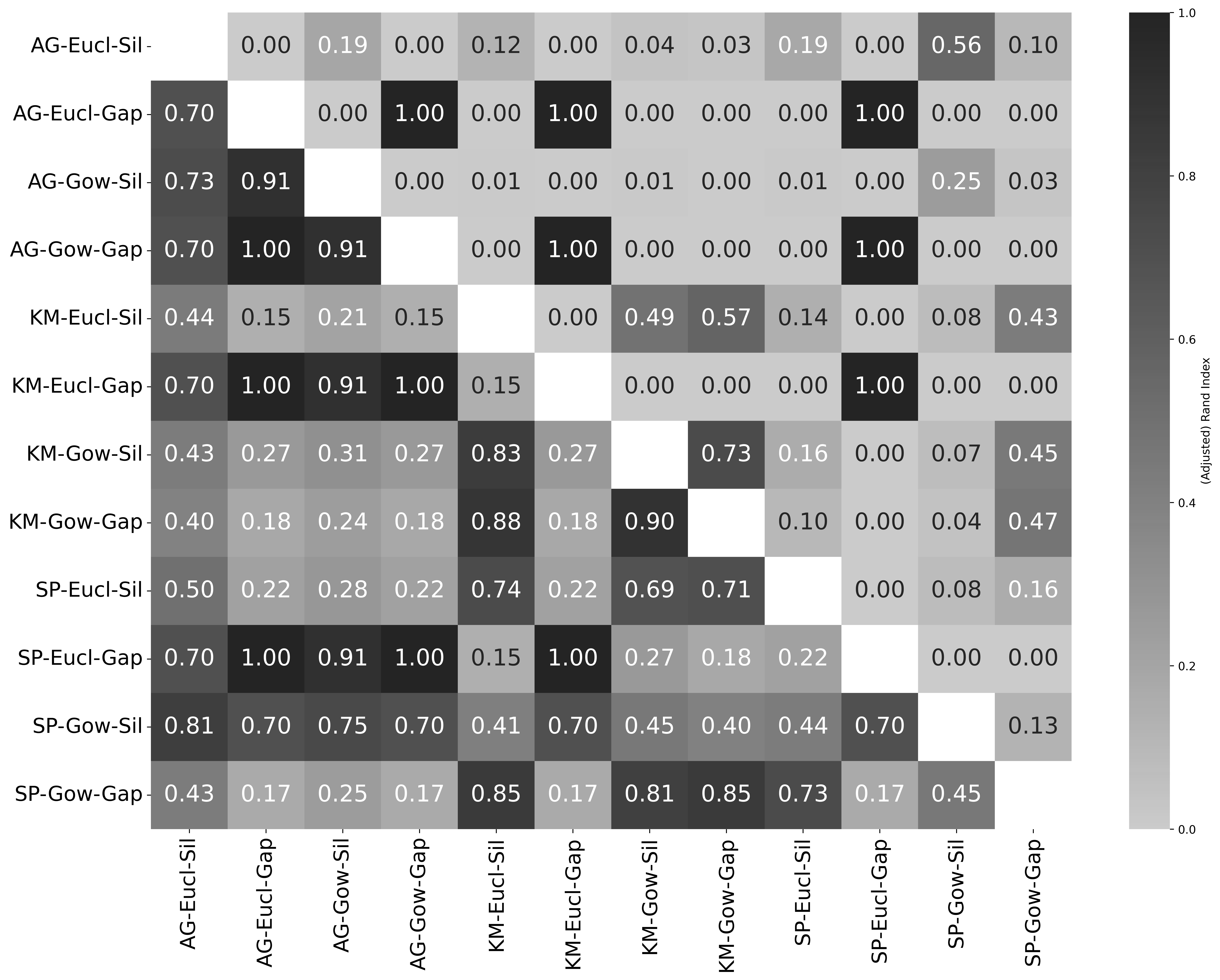} 

}

\caption{Stability between methods. The lower triangular elements of the matrix represent the  Rand indices (RI), while the upper triangular elements show the adjusted Rand indices (ARI).\\\footnotesize{Abbreviations: AG = agglomerative clustering; KM = K-Medoids clustering; SP = spectral clustering; Eucl = Euclidean distance; Gow = Gower's distance; Sil = silhouette value; Gap = gap statistic}}\label{fig:heatmap}
\end{figure}

Stability between methods was low according to the adjusted Rand indices (Figure \ref{fig:heatmap}). Much of the apparent agreement between clustering solutions may be attributed to random overlap rather than meaningful structural similarity. The pairs of strategies showing perfect agreement (i.e., ARI = 1) are combinations of the four strategies with only one cluster. The highest (non-trivial) agreement (ARI 0.73) was found between KM-Gow-Sil (4 clusters) and KM-Gow-Gap (6 clusters), supporting that the silhouette index and gap statistic led to similar clustering for the K-medoids algorithm with Gower's distance in this dataset. In our analysis, none of the clustering solutions had an ARI below zero.

\subsection{Stability of the clustering solutions}\label{stability-of-the-clustering-solutions}

\begin{figure}

{\centering \includegraphics[width=0.9\linewidth]{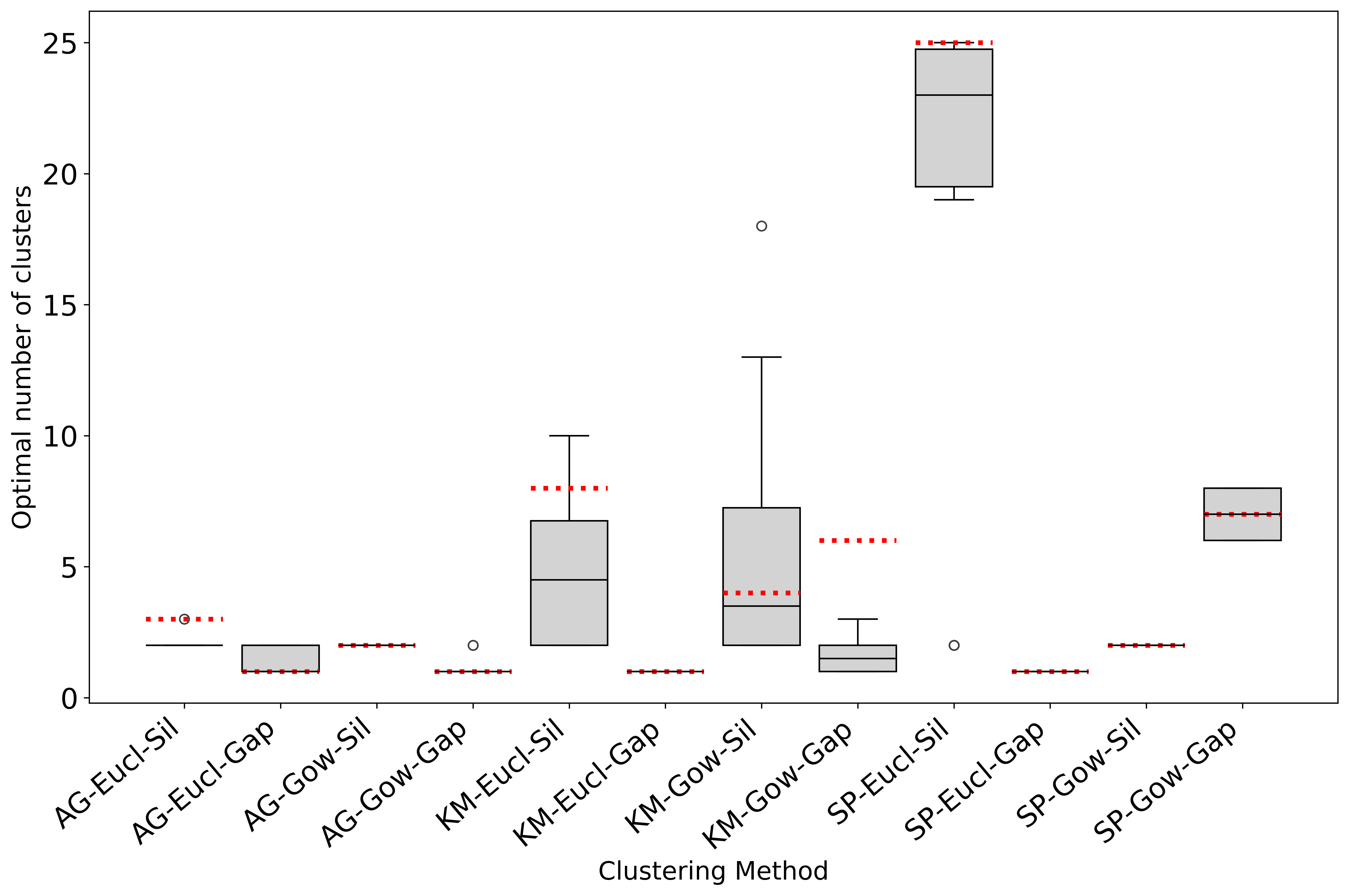} 

}

\caption{Optimal number of clusters identified within bootstrap samples. The red dotted line shows the optimal number of clusters as identified by the clustering algorithm applied on the original dataset.\\\footnotesize{Abbreviations: AG = agglomerative clustering; KM = K-Medoids clustering; SP = spectral clustering; Eucl = Euclidean distance; Gow = Gower's distance; Sil = silhouette value; Gap = gap statistic}}\label{fig:bootstrap}
\end{figure}

To assess the robustness of clustering solutions, we applied all 12 clustering strategies to each of 200 bootstrap datasets. We repeated the full modeling approach in each dataset, such as determining the optimal number of clusters for each bootstrap sample {[}\citeproc{ref-steyerberg2003internal}{26}{]}. Figure \ref{fig:bootstrap} shows the resulting number of clusters per method with the dotted red line indicating the optimal number of clusters identified by the clustering algorithm applied on the original dataset. Notably, the clustering solutions found within the bootstrap samples were highly heterogeneous for most strategies. The number of clusters found for KM-Eucl-Sil, KM-Gow-Sil and SP-Eucl-Sil ranged from two to 10, 13, and 25 respectively. For AG-Eucl-Sil, solutions were limited to either two or three clusters. In contrast, SP-Gow-Sil and AG-Gow-Sil consistently identified two clusters across bootstrap samples. KM-Eucl-Gap and SP-Eucl-Gap, on the other hand, consistently produced a single cluster solution. Interestingly, KM-Gow-Gap identified a six-cluster solution in the original dataset, but this solution never appeared in any of the bootstrap samples.

For each clustering method, we computed all pairwise Rand indices (unadjusted and adjusted) for all \(\frac{200\times199}{2}\) pairs of bootstrap samples. Table \ref{tab:stability} shows the average adjusted and unadjusted Rand indices over all pairs and a summary of the number of clusters and cluster sizes.

The average ARI's show substantial variability across the different methods. For example,
SP-Gow-Sil is stable under data perturbation (average adjusted Rand index is 0.95), while AG-Gow-Sil is relatively unstable (average adjusted Rand index is 0.69). These differences in the adjusted Rand index cannot be explained fully by the different numbers of clusters found for the different methods: both SP-Gow-Sil and AG-Gow-Sil have two clusters, yet their adjusted Rand index differs (0.95 vs 0.69). In contrast, the unadjusted Rand index suggests almost all methods perform similarly well, potentially masking instability. Since the ARI accounts for agreement expected by chance, it highlights meaningful differences that RI alone cannot. This underscores the importance of adjusting for chance before drawing conclusions about clustering stability.

Gap based methods with Euclidean distance (KM-Eucl-Gap and SP-Eucl-Gap) show very high stability with an ARI equal to 1 since these methods frequently resulted in a single cluster solution. When two cluster solutions that assign all observations to a single cluster are compared using ARI, every pair of observations is concordant, yielding an ARI of 1. Their high agreement therefore reflects structural simplicity rather than stable results. SP-Gow-Sil stands out with very high stability (ARI = 0.95) while maintaining a non-degenerate two-cluster structure. AG-Eucl-Sil attains a similarly high ARI (0.88) with three clusters. However, this partition is highly imbalanced, with one cluster containing very few patients and the majority of patients grouped into a single dominant cluster, limiting its practical interpretability.

\subsection{Discriminative ability of the clustering solutions}\label{discriminative-ability-of-the-clustering-solutions}

\begin{figure}

{\centering \includegraphics[width=0.9\linewidth]{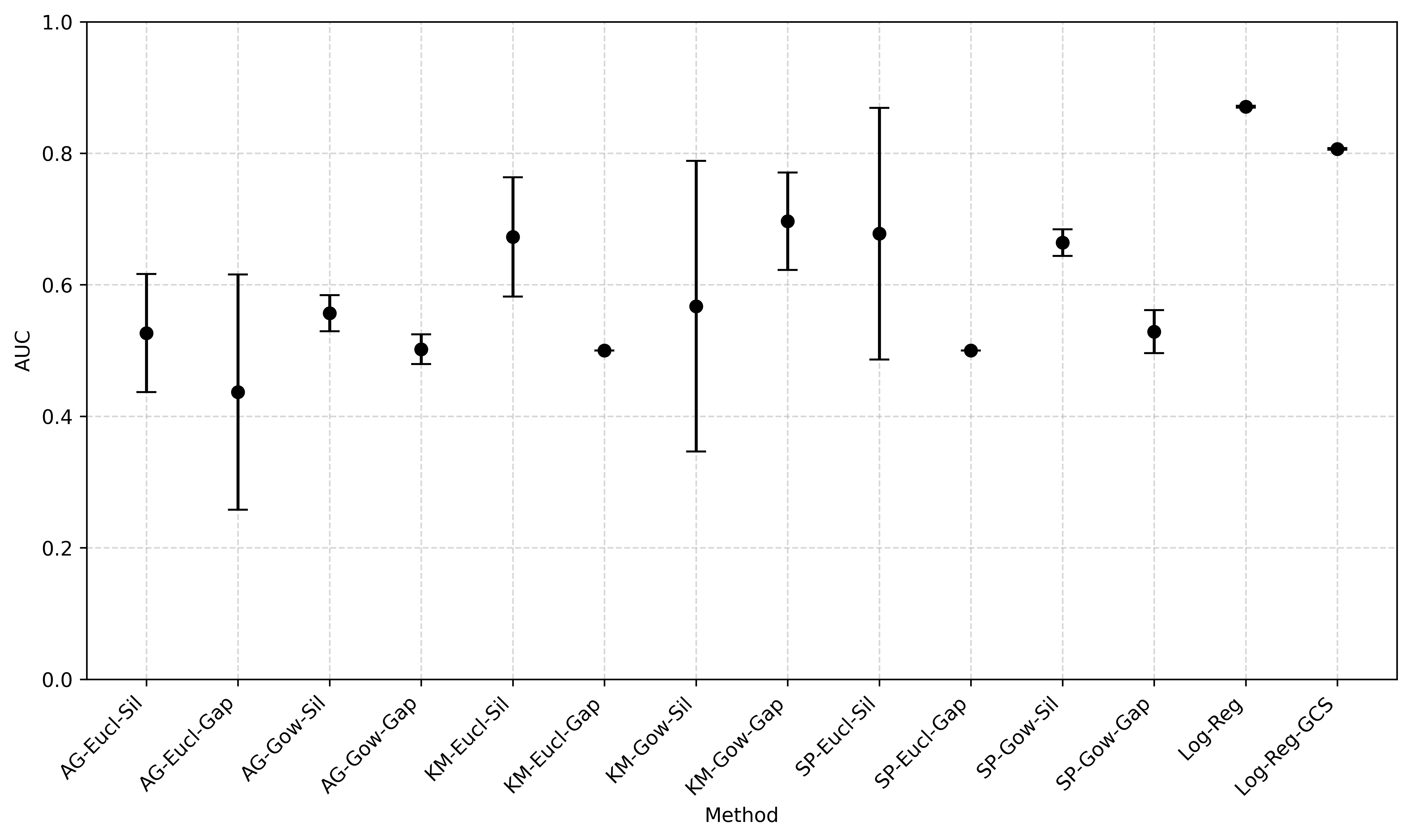} 

}

\caption{Optimism-corrected AUC values across methods with error bars representing the 95\% confidence intervals \\\footnotesize{Abbreviations: AG = agglomerative clustering; KM = K-Medoids clustering; SP = spectral clustering; Eucl = Euclidean distance; Gow = Gower's distance; Sil = silhouette value; Gap = gap statistic; Log-Reg = logistic regression; GCS = Glasgow Coma Scale Score}}\label{fig:auc}
\end{figure}

Notably, the 12 clustering strategies resulted in very different estimates of AUC for predicting the outcome at 6 months (ranging between 0.44 and 0.70), often with wide 95\% confidence intervals indicating instability in predictive performance across bootstrap samples (Figure \ref{fig:auc}). KM-Eucl-Gap and SP-Eucl-Gap had the narrowest 95\% CI's around 0.50. Both these methods consistently identified single cluster solutions during the bootstrap resampling, thus having similar prognostic ability. This narrow confidence intervals reflect methodological degeneracy rather than predictive stability. Among methods producing multi-cluster solutions, AG-Gow-Gap, AG-Gow-Sil and SP-Gow-Sil had the narrowest 95\% CIs ({[}0.48, 0.52{]}, {[}0.53, 0.58{]} and {[}0.64, 0.68{]}, respectively), while KM-Gow-Sil and SP-Eucl-Sil the widest ({[}0.35, 0.79{]} and {[}0.49, 0.87{]}, respectively). This pattern is also evident in the bootstrap distributions of the AUCs which differ greatly in both median and spread of predictive performance between clustering strategies (Supplementary Figure \ref{fig:aucboot}). The width of the confidence intervals depends on the standard deviation of the bootstrap AUCs, therefore, high variability in the AUCs across bootstrap samples leads to wider confidence intervals. Clearly we can see that the clustering strategies that consistently find the similar optimal cluster number during bootstrap also have narrower CI's and the ones that have a wide difference in the optimal cluster size have wider CI's. None of the clustering strategies yield consistently strong prognostic performance. This supports the idea that clustering is not well suited for predicting outcomes in this setting. A standard statistical model like logistic regression, which achieved a stable and high AUC of 0.87 using the same features, is a more appropriate choice for such applications. A logistic regression model including only the GCS score yielded an optimism-adjusted AUC of 0.81, which exceeded the discriminatory performance of all clustering-based approaches. Thus, in this dataset, clustering did not provide added predictive value beyond the GCS score alone.

\section{Discussion}\label{discussion}

In this multiverse-style analysis, the solutions obtained from the different clustering strategies differed substantially in the ``optimal'' number of clusters, the cluster means, and cluster sizes. The differing results obtained with Euclidean and Gower distances show that the apparent structure of patient subgroups is not solely determined by the data itself but is strongly influenced by the choice of distance metric. It highlights the importance of carefully choosing similarity measures in clustering, as they directly influence how patients are grouped. There was little agreement between the strategies as to whether any given pair of patients should be in the same cluster or different clusters. This variation points out that clustering results are highly sensitive to methodological choices in heterogeneous diseases such as TBI. Moreover, clustering solutions were unstable as apparent from variation across bootstrap samples.

Although the relation of patient and disease characteristics to a clinical outcome does not play a role in the optimization criteria of unsupervised clustering approaches, these techniques are applied in clinical research in the hope to identify risk strata in the data. The 12 clustering strategies investigated in this study, showed inconsistent and often limited discriminatory ability and were in this regard clearly inferior to supervised learning methods such as logistic regression. Some solutions produced AUCs with very wide confidence intervals.

Considering that clusters are constructed by optimizing some objective function (determined by the clustering algorithm, distance metric and method to determine the optimal number of clusters), one should not be surprised that different solutions are found for different objective functions. The instability of the clustering solutions found by the same clustering strategy when applied to bootstrap samples of the original data, however, raises questions about whether any such grouping from unsupervised clustering can be interpreted as clinically meaningful. If a clustering solution found in the original data cannot be reproduced in resampled versions of that data, why should we expect this solution to reflect a ``true'' pattern present in the larger patient population?

Our findings contribute to the growing body of literature demonstrating that unsupervised clustering is highly sensitive to analytic decisions. Prior work has shown that different linkage criteria, distance metrics, or graph construction methods can yield different clustering results in the same data {[}\citeproc{ref-steinley2003local}{27}--\citeproc{ref-maier2013result}{29}{]}. Moreover, resampling stability often varies just as much as internal fit indices (optimal number choosing criterion). This suggests that clustering may not be well-suited for risk stratification, where approaches such as supervised learning or model-based methods that explicitly integrate outcomes into subgroup discovery are more appropriate.

A key strength of this study is the systematic variation of the three main decisions in a clustering strategy within a multiverse framework. We used algorithms from three distinct families and selected common choices for the distance metric and cluster selection criteria. This supports the practical relevance of our findings. Investigating within and between strategy stability of the clustering solutions allowed us to evaluate not only the robustness of individual clustering solutions but also the sensitivity of the overall result to analytic decisions.

Several limitations of our study should be acknowledged. We did not include alternatives to the preprocessing steps in our multiverse analysis, such as dimensionality reduction of imaging features, alternative imputation strategies for missing data, or data-driven feature selection--- each of which could have further influenced the clustering results. Even with only varying three components of the clustering strategy, we observe substantial inconsistencies in results, suggesting that further variation would likely lead to even more divergent clustering solutions. That our analysis was restricted to a single dataset may be considered another limitation. While this constrains generalizability, the CENTER-TBI dataset is a landmark resource in traumatic brain injury research, and clustering analyses have been frequently applied in the field {[}\citeproc{ref-tas2024unsupervised}{2}{]}. Given the extent of the variability observed in our study, it is plausible that unsupervised clustering would produce similarly inconsistent results in other clinical applications. Finally, there is a fundamental disconnect between the optimization criteria used in unsupervised clustering and the clinical objective of risk stratification. This issue is not specific to our study but reflects a broader methodological challenge when clustering is used in outcome-related clinical research.

In conclusion, multiple, commonly used and heuristic analytic choices in unsupervised clustering can yield substantially different results. Considering how much the solutions depend on the researcher's decisions and specific set of observations used, it is questionable whether any meaningful interpretations can and should be drawn from unsupervised clustering in clinical research. Therefore, we emphasize that clustering results should not be overinterpreted, as they may reflect artefacts of analytic flexibility rather than meaningful patient subgroups. Researchers should be aware of the limitations of using clustering in risk classification or to support clinical decision making. These techniques seem best applied as exploratory tools, rather than tools for making important disease classifications or as a basis for guiding care.

\section*{Funding Information}\label{funding-information}
\addcontentsline{toc}{section}{Funding Information}

This study was funded by the European Union's Marie Skłodowska-Curie Actions (MSCA) programme under the SOPRANI project (Grant Agreement No.~101119916).

\section*{The SOPRANI collaborators}\label{the-soprani-collaborators}
\addcontentsline{toc}{section}{The SOPRANI collaborators}

Alan Urban, Gabriel Montaldo, Eloïse Baud, Flemish Institute for Biotechnology (VIB), KU Leuven, Leuven, Belgium;

Dick Moberg, Moberg Analytics, Inc., Philadelphia, USA;

Geert Meyfroidt, Fadime Tokmak, Laboratory of Intensive Care Medicine, Department of Cellular and Molecular Medicine, KU Leuven, Leuven, Belgium;

Bart Depreitere, Fabio Gonçalves, and Elle Scheijen, Research Group Experimental Neurosurgery and Neuroanatomy, Department of Neurosciences, KU Leuven, Leuven, Belgium;

Jens P. Dreier, Malaika Mohammad, and Coline L. Lemale, Centre for Stroke Research Berlin, Charité -- Universitätsmedizin Berlin, corporate member of Freie Universität Berlin, Humboldt-Universität zu Berlin, and Berlin Institute of Health, Berlin, Germany;

Samira Saadoun, Marios C. Papadopoulos, and Chibuzor Love Ilochonwu, Department of Neurosurgery, St George's University Hospitals NHS Foundation Trust, London, United Kingdom;

Peter Smielewski and Wenhao Xu, Brain Physics Laboratory, Department of Clinical Neurosciences, University of Cambridge, Cambridge, United Kingdom;

Raimund Helbok and Sara Turella, Department of Neurology, Kepler University Hospital, and Clinical Research Institute for Neuroscience, Johannes Kepler University Linz, Linz, Austria;

Ewout W. Steyerberg, Nicole S. Erler, Eugenia Driusso, and Aneeta Chacko, Julius Center for Health Sciences and Primary Care, University Medical Center Utrecht, Utrecht, The Netherlands;

Wilco Peul, Thomas van Essen, Jeroen van Dijck, and Cansu Rehber, Department of Neurosurgery, University Neurosurgical Center Holland, Leiden University Medical Center, Haaglanden Medical Center, Haga Teaching Hospital, Leiden and The Hague, The Netherlands.

\section*{The CENTER-TBI participants and investigators}\label{the-center-tbi-participants-and-investigators}
\addcontentsline{toc}{section}{The CENTER-TBI participants and investigators}

Cecilia Åkerlund\(^{1}\), Krisztina Amrein\(^{2}\), Nada Andelic\(^{3}\), Lasse Andreassen\(^{4}\), Audny Anke\(^{5}\), Anna Antoni\(^{6}\), Gérard Audibert\(^{7}\), Philippe Azouvi\(^{8}\), Maria Luisa Azzolini\(^{9}\), Ronald Bartels\(^{10}\), Pál Barzó\(^{11}\), Romuald Beauvais\(^{12}\), Ronny Beer\(^{13}\), Bo-Michael Bellander\(^{14}\), Antonio Belli\(^{15}\), Habib Benali\(^{16}\), Maurizio Berardino\(^{17}\), Luigi Beretta\(^{9}\), Morten Blaabjerg\(^{18}\), Peter Bragge\(^{19}\), Alexandra Brazinova\(^{20}\), Vibeke Brinck\(^{21}\), Joanne Brooker\(^{22}\), Camilla Brorsson\(^{23}\), Andras Buki\(^{24}\), Monika Bullinger\(^{25}\), Manuel Cabeleira\(^{26}\), Alessio Caccioppola\(^{27}\), Emiliana Calappi\(^{27}\), Maria Rosa Calvi\(^{9}\), Peter Cameron\(^{28}\), Guillermo Carbayo Lozano\(^{29}\), Marco Carbonara\(^{27}\), Simona Cavallo\(^{17}\), Giorgio Chevallard\(^{30}\), Arturo Chieregato\(^{30}\), Giuseppe Citerio\(^{31, 32}\), Hans Clusmann\(^{33}\), Mark Coburn\(^{34}\), Jonathan Coles\(^{35}\), Jamie D. Cooper\(^{36}\), Marta Correia\(^{37}\), Amra Čović \(^{38}\), Nicola Curry\(^{39}\), Endre Czeiter\(^{40}\), Marek Czosnyka\(^{26}\), Claire Dahyot Fizelier\(^{41}\), Paul Dark\(^{42}\), Helen Dawes\(^{43}\), Véronique De Keyser\(^{44}\), Vincent Degos\(^{16}\), Francesco Della Corte\(^{45}\), Hugo den Boogert\(^{10}\), Bart Depreitere\(^{46}\), Đula Đilvesi\(^{47}\), Abhishek Dixit\(^{48}\), Emma Donoghue\(^{22}\), Jens Dreier\(^{49}\), Guy Loup Dulière\(^{50}\), Ari Ercole\(^{48}\), Patrick Esser\(^{43}\), Erzsébet Ezer\(^{51}\), Martin Fabricius\(^{52}\), Valery L. Feigin\(^{53}\), Kelly Foks\(^{54}\), Shirin Frisvold\(^{55}\), Alex Furmanov\(^{56}\), Pablo Gagliardo\(^{57}\), Damien Galanaud\(^{16}\), Dashiell Gantner\(^{28}\), Guoyi Gao\(^{58}\), Pradeep George\(^{59}\), Alexandre Ghuysen\(^{60}\), Lelde Giga\(^{61}\), Ben Glocker\(^{62}\), Jagoš Golubovic\(^{47}\), Pedro A. Gomez\(^{63}\), Johannes Gratz\(^{64}\), Benjamin Gravesteijn\(^{65}\), Francesca Grossi\(^{45}\), Russell L. Gruen\(^{66}\), Deepak Gupta\(^{67}\), Juanita A. Haagsma\(^{65}\), Iain Haitsma\(^{68}\), Raimund Helbok\(^{69,70}\), Eirik Helseth\(^{71}\), Lindsay Horton \(^{72}\), Jilske Huijben\(^{65}\), Peter J. Hutchinson\(^{73}\), Bram Jacobs\(^{74}\), Stefan Jankowski\(^{75}\), Mike Jarrett\(^{21}\), Ji yao Jiang\(^{59}\), Faye Johnson\(^{76}\), Kelly Jones\(^{53}\), Mladen Karan\(^{47}\), Angelos G. Kolias\(^{73}\), Erwin Kompanje\(^{77}\), Daniel Kondziella\(^{52}\), Evgenios Kornaropoulos\(^{48}\), Lars Owe Koskinen\(^{78}\), Noémi Kovács\(^{79}\), Ana Kowark\(^{80}\), Alfonso Lagares\(^{63}\), Linda Lanyon\(^{59}\), Steven Laureys\(^{81}\), Fiona Lecky\(^{82, 83}\), Didier Ledoux\(^{81}\), Rolf Lefering\(^{84}\), Valerie Legrand\(^{85}\), Aurelie Lejeune\(^{86}\), Leon Levi\(^{87}\), Roger Lightfoot\(^{88}\), Hester Lingsma\(^{65}\), Andrew I.R. Maas\(^{44,89}\), Ana M. Castaño León\(^{63}\), Marc Maegele\(^{90}\), Marek Majdan\(^{20}\), Alex Manara\(^{91}\), Geoffrey Manley\(^{92}\), Costanza Martino\(^{93}\), Hugues Maréchal\(^{50}\), Julia Mattern\(^{94}\), Catherine McMahon\(^{95}\), Béla Melegh\(^{96}\), David Menon\(^{48}\), Tomas Menovsky\(^{44,89}\), Ana Mikolic\(^{65}\), Benoit Misset\(^{81}\), Visakh Muraleedharan\(^{59}\), Lynnette Murray\(^{28}\), Ancuta Negru\(^{97}\), David Nelson\(^{1}\), Virginia Newcombe\(^{48}\), Daan Nieboer\(^{65}\), József Nyirádi\(^{2}\), Otesile Olubukola\(^{82}\), Matej Oresic\(^{98}\), Fabrizio Ortolano\(^{27}\), Aarno Palotie\(^{99, 100, 101}\), Paul M. Parizel\(^{102}\), Jean François Payen\(^{103}\), Natascha Perera\(^{12}\), Vincent Perlbarg\(^{16}\), Paolo Persona\(^{104}\), Wilco Peul\(^{105}\), Anna Piippo-Karjalainen\(^{106}\), Matti Pirinen\(^{99}\), Dana Pisica\(^{65}\), Horia Ples\(^{97}\), Suzanne Polinder\(^{65}\), Inigo Pomposo\(^{29}\), Jussi P. Posti \(^{107}\), Louis Puybasset\(^{108}\), Andreea Radoi\(^{109}\), Arminas Ragauskas\(^{110}\), Rahul Raj\(^{106}\), Malinka Rambadagalla\(^{111}\), Isabel Retel Helmrich\(^{65}\), Jonathan Rhodes\(^{112}\), Sylvia Richardson\(^{113}\), Sophie Richter\(^{48}\), Samuli Ripatti\(^{99}\), Saulius Rocka\(^{110}\), Cecilie Roe\(^{114}\), Olav Roise\(^{115,116}\), Jonathan Rosand\(^{117}\), Jeffrey V. Rosenfeld\(^{118}\), Christina Rosenlund\(^{119}\), Guy Rosenthal\(^{56}\), Rolf Rossaint\(^{80}\), Sandra Rossi\(^{104}\), Daniel Rueckert\(^{62}\), Martin Rusnák\(^{120}\), Juan Sahuquillo\(^{109}\), Oliver Sakowitz\(^{94, 121}\), Renan Sanchez Porras\(^{121}\), Janos Sandor\(^{122}\), Nadine Schäfer\(^{84}\), Silke Schmidt\(^{123}\), Herbert Schoechl\(^{124}\), Guus Schoonman\(^{125}\), Rico Frederik Schou\(^{126}\), Elisabeth Schwendenwein\(^{6}\), Charlie Sewalt\(^{65}\), Ranjit D. Singh\(^{105}\), Toril Skandsen\(^{127}\), 128 , Peter Smielewski\(^{26}\), Abayomi Sorinola\(^{129}\), Emmanuel Stamatakis\(^{48}\), Simon Stanworth\(^{39}\), Robert Stevens\(^{130}\), William Stewart\(^{131}\), Ewout W. Steyerberg\(^{65, 132, 133}\), Nino Stocchetti\(^{134}\), Nina Sundström\(^{135}\), Riikka Takala\(^{136}\), Viktória Tamás\(^{129}\), Tomas Tamosuitis\(^{137}\), Mark Steven Taylor\(^{20}\), Aurore Thibaut\(^{81}\), Braden Te Ao\(^{53}\), Olli Tenovuo\(^{107}\), Alice Theadom\(^{53}\), Matt Thomas\(^{91}\), Dick Tibboel\(^{138}\), Marjolein Timmers\(^{77}\), Christos Tolias\(^{139}\), Tony Trapani\(^{28}\), Cristina Maria Tudora\(^{97}\), Andreas Unterberg\(^{94}\), Peter Vajkoczy\(^{140}\), Shirley Vallance\(^{28}\), Egils Valeinis\(^{61}\), Zoltán Vámos\(^{51}\), Mathieu van der Jagt\(^{141}\), Gregory Van der Steen\(^{44}\), Joukje van der Naalt\(^{74}\), Jeroen T.J.M. van Dijck\(^{105}\), Inge A. M. van Erp\(^{105}\), Thomas A. van Essen\(^{105}\), Wim Van Hecke\(^{142}\), Caroline van Heugten\(^{143}\), Ernest van Veen\(^{65}\), Thijs Vande Vyvere\(^{144}\), Roel P. J. van Wijk\(^{105}\), Alessia Vargiolu\(^{32}\), Emmanuel Vega\(^{86}\), Kimberley Velt\(^{65}\), Jan Verheyden\(^{142}\), Paul M. Vespa\(^{145}\), Anne Vik\(^{127, 146}\), Rimantas Vilcinis\(^{137}\), Victor Volovici\(^{68}\), Nicole von Steinbüchel\(^{38}\), Daphne Voormolen\(^{65}\), Petar Vulekovic\(^{47}\), Kevin K.W. Wang\(^{147}\), Daniel Whitehouse\(^{48}\), Eveline Wiegers\(^{65}\), Guy Williams\(^{48}\), Lindsay Wilson\(^{72}\), Stefan Winzeck\(^{48}\), Stefan Wolf\(^{148}\), Zhihui Yang\(^{117}\), Peter Ylén\(^{149}\), Alexander Younsi\(^{94}\), Marina Zeldovich\(^{150}\), Frederick A. Zeiler\(^{48,151}\), Veronika Zelinkova\(^{20}\), Agate Ziverte\(^{61}\), Tommaso Zoerle\(^{27}\)\\
\textbf{Affliations:}
\(^{1}\) Department of Physiology and Pharmacology, Section of Perioperative Medicine and Intensive Care, Karolinska Institutet, Stockholm, Sweden.\\
\(^{2}\) János Szentágothai Research Centre, University of Pécs, Pécs, Hungary.\\
\(^{3}\) Division of Clinical Neuroscience, Department of Physical Medicine and Rehabilitation, Oslo University Hospital and University of Oslo, Oslo, Norway.\\
\(^{4}\) Department of Neurosurgery, University Hospital Northern Norway, Tromso, Norway.\\
\(^{5}\) Department of Physical Medicine and Rehabilitation, University Hospital Northern Norway, Tromso, Norway.\\
\(^{6}\) Trauma Surgery, Medical University Vienna, Vienna, Austria.\\
\(^{7}\) Department of Anesthesiology \& Intensive Care, University Hospital Nancy, Nancy, France.\\
\(^{8}\) Raymond Poincare hospital, Assistance Publique -- Hopitaux de Paris, Paris, France.\\
\(^{9}\) Department of Anesthesiology \& Intensive Care, S Raffaele University Hospital, Milan, Italy.\\
\(^{10}\) Department of Neurosurgery, Radboud University Medical Center, Nijmegen, The Netherlands.\\
\(^{11}\) Department of Neurosurgery, University of Szeged, Szeged, Hungary.\\
\(^{12}\) International Projects Management, ARTTIC, Munchen, Germany.\\
\(^{13}\) Department of Neurology, Neurological Intensive Care Unit, Medical University of Innsbruck, Innsbruck, Austria.\\
\(^{14}\) Department of Neurosurgery \& Anesthesia \& intensive care medicine, Karolinska University Hospital, Stockholm, Sweden.\\
\(^{15}\) NIHR Surgical Reconstruction and Microbiology Research Centre, Birmingham, UK.\\
\(^{16}\) Anesthesie-Réanimation, Assistance Publique -- Hopitaux de Paris, Paris, France.\\
\(^{17}\) Department of Anesthesia \& ICU, AOU Città della Salute e della Scienza di Torino - Orthopedic and Trauma Center, Torino, Italy.\\
\(^{18}\) Department of Neurology, Odense University Hospital, Odense, Denmark.\\
\(^{19}\) BehaviourWorks Australia, Monash Sustainability Institute, Monash University, Victoria, Australia.\\
\(^{20}\) Department of Public Health, Faculty of Health Sciences and Social Work, Trnava University, Trnava, Slovakia.
\(^{21}\) Quesgen Systems Inc., Burlingame, California, USA.\\
\(^{22}\) Australian \& New Zealand Intensive Care Research Centre, Department of Epidemiology and Preventive Medicine, School of Public Health and Preventive Medicine, Monash University, Melbourne, Australia.\\
\(^{23}\) Department of Surgery and Perioperative Science, Umeå University, Umeå, Sweden.\\
\(^{24}\) Department of Neurosurgery, Örebro University and University Hospital, Örebro, Sweden.\\
\(^{25}\) Department of Medical Psychology, Universitätsklinikum Hamburg-Eppendorf, Hamburg, Germany.\\
\(^{26}\) Brain Physics Lab, Division of Neurosurgery, Dept of Clinical Neurosciences, University of Cambridge, Addenbrooke's Hospital, Cambridge, UK.\\
\(^{27}\) Neuro ICU, Fondazione IRCCS Cà Granda Ospedale Maggiore Policlinico, Milan, Italy.\\
\(^{28}\) ANZIC Research Centre, Monash University, Department of Epidemiology and Preventive Medicine, Melbourne, Victoria, Australia.\\
\(^{29}\) Department of Neurosurgery, Hospital of Cruces, Bilbao, Spain.\\
\(^{30}\) NeuroIntensive Care, Niguarda Hospital, Milan, Italy.\\
\(^{31}\) School of Medicine and Surgery, Università Milano Bicocca, Milano, Italy.\\
\(^{32}\) NeuroIntensive Care Unit, Department Neuroscience, IRCCS Fondazione San Gerardo dei Tintori, Monza, Italy.\\
\(^{33}\) Department of Neurosurgery, Medical Faculty RWTH Aachen University, Aachen, Germany.\\
\(^{34}\) Department of Anesthesiology and Intensive Care Medicine, University Hospital Bonn, Bonn, Germany.\\
\(^{35}\) Department of Anesthesia \& Neurointensive Care, Cambridge University Hospital NHS Foundation Trust, Cambridge, UK.\\
\(^{36}\) School of Public Health \& PM, Monash University and The Alfred Hospital, Melbourne, Victoria, Australia.\\
\(^{37}\) Radiology/MRI department, MRC Cognition and Brain Sciences Unit, Cambridge, UK.\\
\(^{38}\) Institute of Medical Psychology and Medical Sociology, Universitätsmedizin Göttingen, Göttingen, Germany.
\(^{39}\) Oxford University Hospitals NHS Trust, Oxford, UK.\\
\(^{40}\) Department of Neurosurgery, Medical School, University of Pécs, Hungary and Neurotrauma Research Group, János Szentágothai Research Centre, University of Pécs, Hungary.\\
\(^{41}\) Intensive Care Unit, CHU Poitiers, Potiers, France.\\
\(^{42}\) University of Manchester NIHR Biomedical Research Centre, Critical Care Directorate, Salford Royal Hospital NHS Foundation Trust, Salford, UK.\\
\(^{43}\) Movement Science Group, Faculty of Health and Life Sciences, Oxford Brookes University, Oxford, UK.\\
\(^{44}\) Department of Neurosurgery, Antwerp University Hospital, Edegem, Belgium.\\
\(^{45}\) Department of Anesthesia \& Intensive Care, Maggiore Della Carità Hospital, Novara, Italy.\\
\(^{46}\) Department of Neurosurgery, University Hospitals Leuven, Leuven, Belgium.\\
\(^{47}\) Department of Neurosurgery, Clinical centre of Vojvodina, Faculty of Medicine, University of Novi Sad, Novi Sad, Serbia.\\
\(^{48}\) Division of Anaesthesia, University of Cambridge, Addenbrooke's Hospital, Cambridge, UK.\\
\(^{49}\) Center for Stroke Research Berlin, Charité -- Universitätsmedizin Berlin, corporate member of Freie Universität Berlin, Humboldt-Universität zu Berlin, and Berlin Institute of Health, Berlin, Germany.\\
\(^{50}\) Intensive Care Unit, CHR Citadelle, Liège, Belgium.\\
\(^{51}\) Department of Anaesthesiology and Intensive Therapy, University of Pécs, Pécs, Hungary.\\
\(^{52}\) Departments of Neurology, Clinical Neurophysiology and Neuroanesthesiology, Region Hovedstaden Rigshospitalet, Copenhagen, Denmark.\\
\(^{53}\) National Institute for Stroke and Applied Neurosciences, Faculty of Health and Environmental Studies, Auckland University of Technology, Auckland, New Zealand.\\
\(^{54}\) Department of Neurology, Erasmus MC, Rotterdam, the Netherlands.\\
\(^{55}\) Department of Anesthesiology and Intensive care, University Hospital Northern Norway, Tromso, Norway.\\
\(^{56}\) Department of Neurosurgery, Hadassah-hebrew University Medical center, Jerusalem, Israel.\\
\(^{57}\) Fundación Instituto Valenciano de Neurorrehabilitación (FIVAN), Valencia, Spain.\\
\(^{58}\) Department of Neurosurgery, Shanghai Renji hospital, Shanghai Jiaotong University/school of medicine, Shanghai, China.\\
\(^{59}\) Karolinska Institutet, INCF International Neuroinformatics Coordinating Facility, Stockholm, Sweden.\\
\(^{60}\) Emergency Department, CHU, Liège, Belgium.\\
\(^{61}\) Neurosurgery clinic, Pauls Stradins Clinical University Hospital, Riga, Latvia.\\
\(^{62}\) Department of Computing, Imperial College London, London, UK.\\
\(^{63}\) Department of Neurosurgery, Hospital Universitario 12 de Octubre, Madrid, Spain.\\
\(^{64}\) Department of Anesthesia, Critical Care and Pain Medicine, Medical University of Vienna, Austria.\\
\(^{65}\) Department of Public Health, Erasmus Medical Center-University Medical Center, Rotterdam, The Netherlands.\\
\(^{66}\) College of Health and Medicine, Australian National University, Canberra, Australia.\\
\(^{67}\) Department of Neurosurgery, Neurosciences Centre \& JPN Apex trauma centre, All India Institute of Medical Sciences, New Delhi-110029, India.\\
\(^{68}\) Department of Neurosurgery, Erasmus MC, Rotterdam, the Netherlands.\\
\(^{69}\) Department of Neurology, Kepler University Hospital, Johannes Kepler University Linz, Linz, Austria.\\
\(^{70}\) Clinical Research Institute for Neuroscience, Johannes Kepler University Linz, Linz, Austria.\\
\(^{71}\) Department of Neurosurgery, Oslo University Hospital, Oslo, Norway.\\
\(^{72}\) Division of Psychology, University of Stirling, Stirling, UK.\\
\(^{73}\) Division of Neurosurgery, Department of Clinical Neurosciences, Addenbrooke's Hospital \& University of Cambridge, Cambridge, UK.\\
\(^{74}\) Department of Neurology, University of Groningen, University Medical Center Groningen, Groningen, Netherlands.\\
\(^{75}\) Neurointensive Care , Sheffield Teaching Hospitals NHS Foundation Trust, Sheffield, UK.\\
\(^{76}\) Salford Royal Hospital NHS Foundation Trust Acute Research Delivery Team, Salford, UK.\\
\(^{77}\) Department of Intensive Care and Department of Ethics and Philosophy of Medicine, Erasmus Medical Center, Rotterdam, The Netherlands.\\
\(^{78}\) Department of Clinical Neuroscience, Neurosurgery, Umeå University, Umeå, Sweden.\\
\(^{79}\) Hungarian Brain Research Program - Grant No.~KTIA\_13\_NAP-A-II/8, University of Pécs, Pécs, Hungary.\\
\(^{80}\) Department of Anaesthesiology, University Hospital of Aachen, Aachen, Germany.\\
\(^{81}\) Cyclotron Research Center , University of Liège, Liège, Belgium.\\
\(^{82}\) Centre for Urgent and Emergency Care Research (CURE), Health Services Research Section, School of Health and Related Research (ScHARR), University of Sheffield, Sheffield, UK.\\
\(^{83}\) Emergency Department, Salford Royal Hospital, Salford UK.\\
\(^{84}\) Institute of Research in Operative Medicine (IFOM), Witten/Herdecke University, Cologne, Germany.\\
\(^{85}\) VP Global Project Management CNS, ICON, Paris, France.\\
\(^{86}\) Department of Anesthesiology-Intensive Care, Lille University Hospital, Lille, France.\\
\(^{87}\) Department of Neurosurgery, Rambam Medical Center, Haifa, Israel.\\
\(^{88}\) Department of Anesthesiology \& Intensive Care, University Hospitals Southhampton NHS Trust, Southhampton, UK.\\
\(^{89}\) Department of Translational Neuroscience, Faculty of Medicine and Health Science, University of Antwerp, Antwerp, Belgium.\\
\(^{90}\) Cologne-Merheim Medical Center (CMMC), Department of Traumatology, Orthopedic Surgery and Sportmedicine, Witten/Herdecke University, Cologne, Germany.\\
\(^{91}\) Intensive Care Unit, Southmead Hospital, Bristol, Bristol, UK.\\
\(^{92}\) Department of Neurological Surgery, University of California, San Francisco, California, USA.\\
\(^{93}\) Department of Anesthesia \& Intensive Care,M. Bufalini Hospital, Cesena, Italy.\\
\(^{94}\) Department of Neurosurgery, University Hospital Heidelberg, Heidelberg, Germany.\\
\(^{95}\) Department of Neurosurgery, The Walton centre NHS Foundation Trust, Liverpool, UK.\\
\(^{96}\) Department of Medical Genetics, University of Pécs, Pécs, Hungary.\\
\(^{97}\) Department of Neurosurgery, Emergency County Hospital Timisoara , Timisoara, Romania.\\
\(^{98}\) School of Medical Sciences, Örebro University, Örebro, Sweden.\\
\(^{99}\) Institute for Molecular Medicine Finland, University of Helsinki, Helsinki, Finland.\\
\(^{100}\) Analytic and Translational Genetics Unit, Department of Medicine; Psychiatric \& Neurodevelopmental Genetics Unit, Department of Psychiatry; Department of Neurology, Massachusetts General Hospital, Boston, MA, USA.\\
\(^{101}\) Program in Medical and Population Genetics; The Stanley Center for Psychiatric Research, The Broad Institute of MIT and Harvard, Cambridge, MA, USA.\\
\(^{102}\) Department of Radiology, University of Antwerp, Edegem, Belgium.\\
\(^{103}\) Department of Anesthesiology \& Intensive Care, University Hospital of Grenoble, Grenoble, France.\\
\(^{104}\) Department of Anesthesia \& Intensive Care, Azienda Ospedaliera Università di Padova, Padova, Italy.\\
\(^{105}\) Dept. of Neurosurgery, Leiden University Medical Center, Leiden, The Netherlands and Dept. of Neurosurgery, Medical Center Haaglanden, The Hague, The Netherlands.\\
\(^{106}\) Department of Neurosurgery, Helsinki University Central Hospital.\\
\(^{107}\) Division of Clinical Neurosciences, Department of Neurosurgery and Turku Brain Injury Centre, Turku University Hospital and University of Turku, Turku, Finland.\\
\(^{108}\) Department of Anesthesiology and Critical Care, Pitié -Salpêtrière Teaching Hospital, Assistance Publique, Hôpitaux de Paris and University Pierre et Marie Curie, Paris, France.\\
\(^{109}\) Neurotraumatology and Neurosurgery Research Unit (UNINN), Vall d'Hebron Research Institute, Barcelona, Spain.\\
\(^{110}\) Department of Neurosurgery, Kaunas University of technology and Vilnius University, Vilnius, Lithuania.\\
\(^{111}\) Department of Neurosurgery, Rezekne Hospital, Latvia.\\
\(^{112}\) Department of Anaesthesia, Critical Care \& Pain Medicine NHS Lothian \& University of Edinburg, Edinburgh, UK.\\
\(^{113}\) Director, MRC Biostatistics Unit, Cambridge Institute of Public Health, Cambridge, UK.\\
\(^{114}\) Department of Physical Medicine and Rehabilitation, Oslo University Hospital/University of Oslo, Oslo, Norway.\\
\(^{115}\) Division of Orthopedics, Oslo University Hospital, Oslo, Norway.\\
\(^{116}\) Institue of Clinical Medicine, Faculty of Medicine, University of Oslo, Oslo, Norway.\\
\(^{117}\) Broad Institute, Cambridge MA Harvard Medical School, Boston MA, Massachusetts General Hospital, Boston MA, USA.\\
\(^{118}\) National Trauma Research Institute, The Alfred Hospital, Monash University, Melbourne, Victoria, Australia.\\
\(^{119}\) Department of Neurosurgery, Odense University Hospital, Odense, Denmark.\\
\(^{120}\) International Neurotrauma Research Organisation, Vienna, Austria.\\
\(^{121}\) Klinik für Neurochirurgie, Klinikum Ludwigsburg, Ludwigsburg, Germany.\\
\(^{122}\) Division of Biostatistics and Epidemiology, Department of Preventive Medicine, University of Debrecen, Debrecen, Hungary.\\
\(^{123}\) Department Health and Prevention, University Greifswald, Greifswald, Germany.\\
\(^{124}\) Department of Anaesthesiology and Intensive Care, AUVA Trauma Hospital, Salzburg, Austria.\\
\(^{125}\) Department of Neurology, Elisabeth-TweeSteden Ziekenhuis, Tilburg, the Netherlands.\\
\(^{126}\) Department of Neuroanesthesia and Neurointensive Care, Odense University Hospital, Odense, Denmark.\\
\(^{127}\) Department of Neuromedicine and Movement Science, Norwegian University of Science and Technology, NTNU, Trondheim, Norway.\\
\(^{128}\) Department of Physical Medicine and Rehabilitation, St.Olavs Hospital, Trondheim University Hospital, Trondheim, Norway.\\
\(^{129}\) Department of Neurosurgery, University of Pécs, Pécs, Hungary.\\
\(^{130}\) Division of Neuroscience Critical Care, John Hopkins University School of Medicine, Baltimore, USA.\\
\(^{131}\) Department of Neuropathology, Queen Elizabeth University Hospital and University of Glasgow, Glasgow, UK.\\
\(^{132}\) Dept. of Department of Biomedical Data Sciences, Leiden University Medical Center, Leiden, The Netherlands.\\
\(^{133}\) Julius Center for Health Sciences and Primary Care, University Medical Center Utrecht, Utrecht, The Netherlands.\\
\(^{134}\) Department of Pathophysiology and Transplantation, Milan University, and Neuroscience ICU, Fondazione IRCCS Cà Granda Ospedale Maggiore Policlinico, Milano, Italy.\\
\(^{135}\) Department of Radiation Sciences, Biomedical Engineering, Umeå University, Umeå, Sweden.\\
\(^{136}\) Perioperative Services, Intensive Care Medicine and Pain Management, Turku University Hospital and University of Turku, Turku, Finland.\\
\(^{137}\) Department of Neurosurgery, Kaunas University of Health Sciences, Kaunas, Lithuania.\\
\(^{138}\) Intensive Care and Department of Pediatric Surgery, Erasmus Medical Center, Sophia Children's Hospital, Rotterdam, The Netherlands.\\
\(^{139}\) Department of Neurosurgery, Kings college London, London, UK.\\
\(^{140}\) Neurologie, Neurochirurgie und Psychiatrie, Charité -- Universitätsmedizin Berlin, Berlin, Germany.\\
\(^{141}\) Department of Intensive Care Adults, Erasmus MC-- University Medical Center Rotterdam, Rotterdam, the Netherlands.\\
\(^{142}\) icoMetrix NV, Leuven, Belgium.\\
\(^{143}\) Movement Science Group, Faculty of Health and Life Sciences, Oxford Brookes University, Oxford, UK.\\
\(^{144}\) Radiology Department, Antwerp University Hospital and University of Antwerp, Edegem, Belgium.\\
\(^{145}\) Director of Neurocritical Care, University of California, Los Angeles, USA.\\
\(^{146}\) Department of Neurosurgery, St.Olavs Hospital, Trondheim University Hospital, Trondheim, Norway.\\
\(^{147}\) Department of Emergency Medicine, University of Florida, Gainesville, Florida, USA.\\
\(^{148}\) Department of Neurosurgery, Charité -- Universitätsmedizin Berlin, corporate member of Freie Universität Berlin, Humboldt-Universität zu Berlin, and Berlin Institute of Health, Berlin, Germany.\\
\(^{149}\) VTT Technical Research Centre, Tampere, Finland.\\
\(^{150}\) Sigmund Freud University, Faculty of Psychotherapy Science, Vienna, Austria.\\
\(^{151}\) Section of Neurosurgery, Department of Surgery, Rady Faculty of Health Sciences, University of Manitoba, Winnipeg, MB, Canada.

\section*{References}\label{references}
\addcontentsline{toc}{section}{References}

\protect\phantomsection\label{refs}

\clearpage

\section*{Tables}\label{tables}
\addcontentsline{toc}{section}{Tables}

\global\setlength{\Oldarrayrulewidth}{\arrayrulewidth}

\global\setlength{\Oldtabcolsep}{\tabcolsep}

\setlength{\tabcolsep}{2pt}

\renewcommand*{\arraystretch}{1.5}

\providecommand{\ascline}[3]{\noalign{\global\arrayrulewidth #1}\arrayrulecolor[HTML]{#2}\cline{#3}}

\begin{longtable}[c]{|p{2.86in}|p{1.12in}|p{1.08in}}

\caption{Overview\ of\ the\ CENTER-TBI\ data.}\label{tab:tab1}\\

\ascline{1.5pt}{666666}{1-3}

\multicolumn{1}{>{\raggedright}m{\dimexpr 2.86in+0\tabcolsep}}{\textcolor[HTML]{000000}{\fontsize{10}{10}\selectfont{{Baseline\ Characteristics}}}} & \multicolumn{1}{>{\centering}m{\dimexpr 1.12in+0\tabcolsep}}{\textcolor[HTML]{000000}{\fontsize{10}{10}\selectfont{{n\ =\ 4,509}}}} & \multicolumn{1}{>{\centering}m{\dimexpr 1.08in+0\tabcolsep}}{\textcolor[HTML]{000000}{\fontsize{10}{10}\selectfont{{Missing\ (\%)}}}} \\

\ascline{1.5pt}{666666}{1-3}\endfirsthead \caption[]{Overview\ of\ the\ CENTER-TBI\ data.}\label{tab:tab1}\\

\ascline{1.5pt}{666666}{1-3}

\multicolumn{1}{>{\raggedright}m{\dimexpr 2.86in+0\tabcolsep}}{\textcolor[HTML]{000000}{\fontsize{10}{10}\selectfont{{Baseline\ Characteristics}}}} & \multicolumn{1}{>{\centering}m{\dimexpr 1.12in+0\tabcolsep}}{\textcolor[HTML]{000000}{\fontsize{10}{10}\selectfont{{n\ =\ 4,509}}}} & \multicolumn{1}{>{\centering}m{\dimexpr 1.08in+0\tabcolsep}}{\textcolor[HTML]{000000}{\fontsize{10}{10}\selectfont{{Missing\ (\%)}}}} \\

\ascline{1.5pt}{666666}{1-3}\endhead

\multicolumn{1}{>{\raggedright}m{\dimexpr 2.86in+0\tabcolsep}}{\textcolor[HTML]{000000}{\fontsize{10}{10}\selectfont{{Age\ (Median\ [IQR])}}}} & \multicolumn{1}{>{\centering}m{\dimexpr 1.12in+0\tabcolsep}}{\textcolor[HTML]{000000}{\fontsize{10}{10}\selectfont{{50\ [30\ -\ 66]}}}} & \multicolumn{1}{>{\centering}m{\dimexpr 1.08in+0\tabcolsep}}{\textcolor[HTML]{000000}{\fontsize{10}{10}\selectfont{{0.0}}}} \\

\multicolumn{1}{>{\raggedright}m{\dimexpr 2.86in+0\tabcolsep}}{\textcolor[HTML]{000000}{\fontsize{10}{10}\selectfont{{Injury\ Cause}}}} & \multicolumn{1}{>{\centering}m{\dimexpr 1.12in+0\tabcolsep}}{\textcolor[HTML]{000000}{\fontsize{10}{10}\selectfont{{}}}} & \multicolumn{1}{>{\centering}m{\dimexpr 1.08in+0\tabcolsep}}{\textcolor[HTML]{000000}{\fontsize{10}{10}\selectfont{{3.7}}}} \\

\multicolumn{1}{>{\raggedright}m{\dimexpr 2.86in+0\tabcolsep}}{\textcolor[HTML]{000000}{\fontsize{10}{10}\selectfont{{\ \ Road\ traffic\ accident}}}} & \multicolumn{1}{>{\centering}m{\dimexpr 1.12in+0\tabcolsep}}{\textcolor[HTML]{000000}{\fontsize{10}{10}\selectfont{{1,682\ (39\%)}}}} & \multicolumn{1}{>{\centering}m{\dimexpr 1.08in+0\tabcolsep}}{\textcolor[HTML]{000000}{\fontsize{10}{10}\selectfont{{}}}} \\

\multicolumn{1}{>{\raggedright}m{\dimexpr 2.86in+0\tabcolsep}}{\textcolor[HTML]{000000}{\fontsize{10}{10}\selectfont{{\ \ Fall}}}} & \multicolumn{1}{>{\centering}m{\dimexpr 1.12in+0\tabcolsep}}{\textcolor[HTML]{000000}{\fontsize{10}{10}\selectfont{{2,024\ (47\%)}}}} & \multicolumn{1}{>{\centering}m{\dimexpr 1.08in+0\tabcolsep}}{\textcolor[HTML]{000000}{\fontsize{10}{10}\selectfont{{}}}} \\

\multicolumn{1}{>{\raggedright}m{\dimexpr 2.86in+0\tabcolsep}}{\textcolor[HTML]{000000}{\fontsize{10}{10}\selectfont{{\ \ Violence/Suicide}}}} & \multicolumn{1}{>{\centering}m{\dimexpr 1.12in+0\tabcolsep}}{\textcolor[HTML]{000000}{\fontsize{10}{10}\selectfont{{293\ (6.7\%)}}}} & \multicolumn{1}{>{\centering}m{\dimexpr 1.08in+0\tabcolsep}}{\textcolor[HTML]{000000}{\fontsize{10}{10}\selectfont{{}}}} \\

\multicolumn{1}{>{\raggedright}m{\dimexpr 2.86in+0\tabcolsep}}{\textcolor[HTML]{000000}{\fontsize{10}{10}\selectfont{{\ \ Other}}}} & \multicolumn{1}{>{\centering}m{\dimexpr 1.12in+0\tabcolsep}}{\textcolor[HTML]{000000}{\fontsize{10}{10}\selectfont{{343\ (7.9\%)}}}} & \multicolumn{1}{>{\centering}m{\dimexpr 1.08in+0\tabcolsep}}{\textcolor[HTML]{000000}{\fontsize{10}{10}\selectfont{{}}}} \\

\multicolumn{1}{>{\raggedright}m{\dimexpr 2.86in+0\tabcolsep}}{\textcolor[HTML]{000000}{\fontsize{10}{10}\selectfont{{GCS\ Motor\ (Median\ [IQR])}}}} & \multicolumn{1}{>{\centering}m{\dimexpr 1.12in+0\tabcolsep}}{\textcolor[HTML]{000000}{\fontsize{10}{10}\selectfont{{6\ [5\ -\ 6]}}}} & \multicolumn{1}{>{\centering}m{\dimexpr 1.08in+0\tabcolsep}}{\textcolor[HTML]{000000}{\fontsize{10}{10}\selectfont{{2.4}}}} \\

\multicolumn{1}{>{\raggedright}m{\dimexpr 2.86in+0\tabcolsep}}{\textcolor[HTML]{000000}{\fontsize{10}{10}\selectfont{{GCS\ sum\ Score\ (Median\ [IQR])}}}} & \multicolumn{1}{>{\centering}m{\dimexpr 1.12in+0\tabcolsep}}{\textcolor[HTML]{000000}{\fontsize{10}{10}\selectfont{{15\ [10\ -\ 15]}}}} & \multicolumn{1}{>{\centering}m{\dimexpr 1.08in+0\tabcolsep}}{\textcolor[HTML]{000000}{\fontsize{10}{10}\selectfont{{3.9}}}} \\

\multicolumn{1}{>{\raggedright}m{\dimexpr 2.86in+0\tabcolsep}}{\textcolor[HTML]{000000}{\fontsize{10}{10}\selectfont{{Pupillary\ Reactivity}}}} & \multicolumn{1}{>{\centering}m{\dimexpr 1.12in+0\tabcolsep}}{\textcolor[HTML]{000000}{\fontsize{10}{10}\selectfont{{}}}} & \multicolumn{1}{>{\centering}m{\dimexpr 1.08in+0\tabcolsep}}{\textcolor[HTML]{000000}{\fontsize{10}{10}\selectfont{{5.8}}}} \\

\multicolumn{1}{>{\raggedright}m{\dimexpr 2.86in+0\tabcolsep}}{\textcolor[HTML]{000000}{\fontsize{10}{10}\selectfont{{\ \ Both\ reactive}}}} & \multicolumn{1}{>{\centering}m{\dimexpr 1.12in+0\tabcolsep}}{\textcolor[HTML]{000000}{\fontsize{10}{10}\selectfont{{3,802\ (90\%)}}}} & \multicolumn{1}{>{\centering}m{\dimexpr 1.08in+0\tabcolsep}}{\textcolor[HTML]{000000}{\fontsize{10}{10}\selectfont{{}}}} \\

\multicolumn{1}{>{\raggedright}m{\dimexpr 2.86in+0\tabcolsep}}{\textcolor[HTML]{000000}{\fontsize{10}{10}\selectfont{{\ \ One\ reactive}}}} & \multicolumn{1}{>{\centering}m{\dimexpr 1.12in+0\tabcolsep}}{\textcolor[HTML]{000000}{\fontsize{10}{10}\selectfont{{164\ (3.9\%)}}}} & \multicolumn{1}{>{\centering}m{\dimexpr 1.08in+0\tabcolsep}}{\textcolor[HTML]{000000}{\fontsize{10}{10}\selectfont{{}}}} \\

\multicolumn{1}{>{\raggedright}m{\dimexpr 2.86in+0\tabcolsep}}{\textcolor[HTML]{000000}{\fontsize{10}{10}\selectfont{{\ \ None\ reactive}}}} & \multicolumn{1}{>{\centering}m{\dimexpr 1.12in+0\tabcolsep}}{\textcolor[HTML]{000000}{\fontsize{10}{10}\selectfont{{281\ (6.6\%)}}}} & \multicolumn{1}{>{\centering}m{\dimexpr 1.08in+0\tabcolsep}}{\textcolor[HTML]{000000}{\fontsize{10}{10}\selectfont{{}}}} \\

\multicolumn{1}{>{\raggedright}m{\dimexpr 2.86in+0\tabcolsep}}{\textcolor[HTML]{000000}{\fontsize{10}{10}\selectfont{{Pre-admission\ hypoxia}}}} & \multicolumn{1}{>{\centering}m{\dimexpr 1.12in+0\tabcolsep}}{\textcolor[HTML]{000000}{\fontsize{10}{10}\selectfont{{299\ (7.0\%)}}}} & \multicolumn{1}{>{\centering}m{\dimexpr 1.08in+0\tabcolsep}}{\textcolor[HTML]{000000}{\fontsize{10}{10}\selectfont{{5.6}}}} \\

\multicolumn{1}{>{\raggedright}m{\dimexpr 2.86in+0\tabcolsep}}{\textcolor[HTML]{000000}{\fontsize{10}{10}\selectfont{{Pre-admission\ hypotension}}}} & \multicolumn{1}{>{\centering}m{\dimexpr 1.12in+0\tabcolsep}}{\textcolor[HTML]{000000}{\fontsize{10}{10}\selectfont{{297\ (6.9\%)}}}} & \multicolumn{1}{>{\centering}m{\dimexpr 1.08in+0\tabcolsep}}{\textcolor[HTML]{000000}{\fontsize{10}{10}\selectfont{{4.7}}}} \\

\multicolumn{1}{>{\raggedright}m{\dimexpr 2.86in+0\tabcolsep}}{\textcolor[HTML]{000000}{\fontsize{10}{10}\selectfont{{Major\ Extracranial\ Injury}}}\textcolor[HTML]{000000}{\fontsize{10}{10}\selectfont{{\textsuperscript{a}}}}} & \multicolumn{1}{>{\centering}m{\dimexpr 1.12in+0\tabcolsep}}{\textcolor[HTML]{000000}{\fontsize{10}{10}\selectfont{{668\ (15\%)}}}} & \multicolumn{1}{>{\centering}m{\dimexpr 1.08in+0\tabcolsep}}{\textcolor[HTML]{000000}{\fontsize{10}{10}\selectfont{{0.0}}}} \\

\multicolumn{1}{>{\raggedright}m{\dimexpr 2.86in+0\tabcolsep}}{\textcolor[HTML]{000000}{\fontsize{10}{10}\selectfont{{Patient\ Stratum}}}} & \multicolumn{1}{>{\centering}m{\dimexpr 1.12in+0\tabcolsep}}{\textcolor[HTML]{000000}{\fontsize{10}{10}\selectfont{{}}}} & \multicolumn{1}{>{\centering}m{\dimexpr 1.08in+0\tabcolsep}}{\textcolor[HTML]{000000}{\fontsize{10}{10}\selectfont{{0.0}}}} \\

\multicolumn{1}{>{\raggedright}m{\dimexpr 2.86in+0\tabcolsep}}{\textcolor[HTML]{000000}{\fontsize{10}{10}\selectfont{{\ \ ER}}}} & \multicolumn{1}{>{\centering}m{\dimexpr 1.12in+0\tabcolsep}}{\textcolor[HTML]{000000}{\fontsize{10}{10}\selectfont{{848\ (19\%)}}}} & \multicolumn{1}{>{\centering}m{\dimexpr 1.08in+0\tabcolsep}}{\textcolor[HTML]{000000}{\fontsize{10}{10}\selectfont{{}}}} \\

\multicolumn{1}{>{\raggedright}m{\dimexpr 2.86in+0\tabcolsep}}{\textcolor[HTML]{000000}{\fontsize{10}{10}\selectfont{{\ \ Admission}}}} & \multicolumn{1}{>{\centering}m{\dimexpr 1.12in+0\tabcolsep}}{\textcolor[HTML]{000000}{\fontsize{10}{10}\selectfont{{1,523\ (34\%)}}}} & \multicolumn{1}{>{\centering}m{\dimexpr 1.08in+0\tabcolsep}}{\textcolor[HTML]{000000}{\fontsize{10}{10}\selectfont{{}}}} \\

\multicolumn{1}{>{\raggedright}m{\dimexpr 2.86in+0\tabcolsep}}{\textcolor[HTML]{000000}{\fontsize{10}{10}\selectfont{{\ \ ICU}}}} & \multicolumn{1}{>{\centering}m{\dimexpr 1.12in+0\tabcolsep}}{\textcolor[HTML]{000000}{\fontsize{10}{10}\selectfont{{2,138\ (47\%)}}}} & \multicolumn{1}{>{\centering}m{\dimexpr 1.08in+0\tabcolsep}}{\textcolor[HTML]{000000}{\fontsize{10}{10}\selectfont{{}}}} \\

\multicolumn{1}{>{\raggedright}m{\dimexpr 2.86in+0\tabcolsep}}{\textcolor[HTML]{000000}{\fontsize{10}{10}\selectfont{{Axonal\ injury}}}} & \multicolumn{1}{>{\centering}m{\dimexpr 1.12in+0\tabcolsep}}{\textcolor[HTML]{000000}{\fontsize{10}{10}\selectfont{{401\ (9.3\%)}}}} & \multicolumn{1}{>{\centering}m{\dimexpr 1.08in+0\tabcolsep}}{\textcolor[HTML]{000000}{\fontsize{10}{10}\selectfont{{14}}}} \\

\multicolumn{1}{>{\raggedright}m{\dimexpr 2.86in+0\tabcolsep}}{\textcolor[HTML]{000000}{\fontsize{10}{10}\selectfont{{Contusion}}}} & \multicolumn{1}{>{\centering}m{\dimexpr 1.12in+0\tabcolsep}}{\textcolor[HTML]{000000}{\fontsize{10}{10}\selectfont{{1,097\ (25\%)}}}} & \multicolumn{1}{>{\centering}m{\dimexpr 1.08in+0\tabcolsep}}{\textcolor[HTML]{000000}{\fontsize{10}{10}\selectfont{{14}}}} \\

\multicolumn{1}{>{\raggedright}m{\dimexpr 2.86in+0\tabcolsep}}{\textcolor[HTML]{000000}{\fontsize{10}{10}\selectfont{{Subdural\ hematoma\ subacute\ chronic}}}} & \multicolumn{1}{>{\centering}m{\dimexpr 1.12in+0\tabcolsep}}{\textcolor[HTML]{000000}{\fontsize{10}{10}\selectfont{{82\ (1.9\%)}}}} & \multicolumn{1}{>{\centering}m{\dimexpr 1.08in+0\tabcolsep}}{\textcolor[HTML]{000000}{\fontsize{10}{10}\selectfont{{14}}}} \\

\multicolumn{1}{>{\raggedright}m{\dimexpr 2.86in+0\tabcolsep}}{\textcolor[HTML]{000000}{\fontsize{10}{10}\selectfont{{Traumatic\ SAH}}}} & \multicolumn{1}{>{\centering}m{\dimexpr 1.12in+0\tabcolsep}}{\textcolor[HTML]{000000}{\fontsize{10}{10}\selectfont{{1,469\ (34\%)}}}} & \multicolumn{1}{>{\centering}m{\dimexpr 1.08in+0\tabcolsep}}{\textcolor[HTML]{000000}{\fontsize{10}{10}\selectfont{{14}}}} \\

\multicolumn{1}{>{\raggedright}m{\dimexpr 2.86in+0\tabcolsep}}{\textcolor[HTML]{000000}{\fontsize{10}{10}\selectfont{{Epidural\ hematoma}}}} & \multicolumn{1}{>{\centering}m{\dimexpr 1.12in+0\tabcolsep}}{\textcolor[HTML]{000000}{\fontsize{10}{10}\selectfont{{349\ (8.0\%)}}}} & \multicolumn{1}{>{\centering}m{\dimexpr 1.08in+0\tabcolsep}}{\textcolor[HTML]{000000}{\fontsize{10}{10}\selectfont{{14}}}} \\

\multicolumn{1}{>{\raggedright}m{\dimexpr 2.86in+0\tabcolsep}}{\textcolor[HTML]{000000}{\fontsize{10}{10}\selectfont{{Subdural\ hematoma\ acute}}}} & \multicolumn{1}{>{\centering}m{\dimexpr 1.12in+0\tabcolsep}}{\textcolor[HTML]{000000}{\fontsize{10}{10}\selectfont{{889\ (21\%)}}}} & \multicolumn{1}{>{\centering}m{\dimexpr 1.08in+0\tabcolsep}}{\textcolor[HTML]{000000}{\fontsize{10}{10}\selectfont{{14}}}} \\

\multicolumn{1}{>{\raggedright}m{\dimexpr 2.86in+0\tabcolsep}}{\textcolor[HTML]{000000}{\fontsize{10}{10}\selectfont{{Skull\ fracture}}}} & \multicolumn{1}{>{\centering}m{\dimexpr 1.12in+0\tabcolsep}}{\textcolor[HTML]{000000}{\fontsize{10}{10}\selectfont{{1,156\ (27\%)}}}} & \multicolumn{1}{>{\centering}m{\dimexpr 1.08in+0\tabcolsep}}{\textcolor[HTML]{000000}{\fontsize{10}{10}\selectfont{{14}}}} \\

\multicolumn{1}{>{\raggedright}m{\dimexpr 2.86in+0\tabcolsep}}{\textcolor[HTML]{000000}{\fontsize{10}{10}\selectfont{{Subdural\ collection\ mixed\ density}}}} & \multicolumn{1}{>{\centering}m{\dimexpr 1.12in+0\tabcolsep}}{\textcolor[HTML]{000000}{\fontsize{10}{10}\selectfont{{69\ (1.6\%)}}}} & \multicolumn{1}{>{\centering}m{\dimexpr 1.08in+0\tabcolsep}}{\textcolor[HTML]{000000}{\fontsize{10}{10}\selectfont{{14}}}} \\

\multicolumn{1}{>{\raggedright}m{\dimexpr 2.86in+0\tabcolsep}}{\textcolor[HTML]{000000}{\fontsize{10}{10}\selectfont{{Cisternal\ compression}}}} & \multicolumn{1}{>{\centering}m{\dimexpr 1.12in+0\tabcolsep}}{\textcolor[HTML]{000000}{\fontsize{10}{10}\selectfont{{480\ (11\%)}}}} & \multicolumn{1}{>{\centering}m{\dimexpr 1.08in+0\tabcolsep}}{\textcolor[HTML]{000000}{\fontsize{10}{10}\selectfont{{14}}}} \\

\multicolumn{1}{>{\raggedright}m{\dimexpr 2.86in+0\tabcolsep}}{\textcolor[HTML]{000000}{\fontsize{10}{10}\selectfont{{Midline\ shift}}}} & \multicolumn{1}{>{\centering}m{\dimexpr 1.12in+0\tabcolsep}}{\textcolor[HTML]{000000}{\fontsize{10}{10}\selectfont{{349\ (8.1\%)}}}} & \multicolumn{1}{>{\centering}m{\dimexpr 1.08in+0\tabcolsep}}{\textcolor[HTML]{000000}{\fontsize{10}{10}\selectfont{{14}}}} \\

\multicolumn{1}{>{\raggedright}m{\dimexpr 2.86in+0\tabcolsep}}{\textcolor[HTML]{000000}{\fontsize{10}{10}\selectfont{{Mass\ lesion}}}} & \multicolumn{1}{>{\centering}m{\dimexpr 1.12in+0\tabcolsep}}{\textcolor[HTML]{000000}{\fontsize{10}{10}\selectfont{{546\ (13\%)}}}} & \multicolumn{1}{>{\centering}m{\dimexpr 1.08in+0\tabcolsep}}{\textcolor[HTML]{000000}{\fontsize{10}{10}\selectfont{{14}}}} \\

\multicolumn{1}{>{\raggedright}m{\dimexpr 2.86in+0\tabcolsep}}{\textcolor[HTML]{000000}{\fontsize{10}{10}\selectfont{{Intraventricular\ hemorrhage}}}} & \multicolumn{1}{>{\centering}m{\dimexpr 1.12in+0\tabcolsep}}{\textcolor[HTML]{000000}{\fontsize{10}{10}\selectfont{{408\ (9.4\%)}}}} & \multicolumn{1}{>{\centering}m{\dimexpr 1.08in+0\tabcolsep}}{\textcolor[HTML]{000000}{\fontsize{10}{10}\selectfont{{14}}}} \\

\ascline{1.5pt}{666666}{1-3}

\end{longtable}

\arrayrulecolor[HTML]{000000}

\global\setlength{\arrayrulewidth}{\Oldarrayrulewidth}

\global\setlength{\tabcolsep}{\Oldtabcolsep}

\renewcommand*{\arraystretch}{1}

\begin{footnotesize} Note: Median and IQR are shown for numeric values (IQR = 75th percentile – 25th percentile), while count and percentage (of the observed cases) are shown for categorical variables. GCS Sum Score and Patient Stratum were used in the imputation but were not included in the clustering analysis. \textsuperscript{a}Defined as non-head Abbreviated Injury Scale (AIS) $\ge 3$. Abbreviations: IQR = Interquartile range; GCS = Glasgow Coma Scale; ER = Emergency Room; ICU = Intensive Care Unit; SAH = Subarachnoid Hemorrhage \end{footnotesize}

\global\setlength{\Oldarrayrulewidth}{\arrayrulewidth}

\global\setlength{\Oldtabcolsep}{\tabcolsep}

\setlength{\tabcolsep}{2pt}

\renewcommand*{\arraystretch}{1.5}

\providecommand{\ascline}[3]{\noalign{\global\arrayrulewidth #1}\arrayrulecolor[HTML]{#2}\cline{#3}}

\begin{longtable}[c]{|p{1.50in}|p{1.50in}|p{1.50in}|p{1.50in}}

\caption{Within\ method\ stability\ indices\ obtained\ from\ the\ bootstrap\ samples}\label{tab:stability}\\

\ascline{1.5pt}{666666}{1-4}

\multicolumn{1}{>{\raggedright}m{\dimexpr 1.5in+0\tabcolsep}}{\textcolor[HTML]{000000}{\fontsize{10}{10}\selectfont{{Algorithm}}}} & \multicolumn{1}{>{\centering}m{\dimexpr 1.5in+0\tabcolsep}}{\textcolor[HTML]{000000}{\fontsize{10}{10}\selectfont{{Range\ of\ the\ optimal\ number\ of\ clusters\ found\ in\ the\ bootstrap\ samples}}}} & \multicolumn{1}{>{\centering}m{\dimexpr 1.5in+0\tabcolsep}}{\textcolor[HTML]{000000}{\fontsize{10}{10}\selectfont{{Rand\ Index\ [0,\ 1]}}}} & \multicolumn{1}{>{\centering}m{\dimexpr 1.5in+0\tabcolsep}}{\textcolor[HTML]{000000}{\fontsize{10}{10}\selectfont{{Adjusted\ Rand\ Index\ [-1,\ 1]}}}} \\

\ascline{1.5pt}{666666}{1-4}\endfirsthead \caption[]{Within\ method\ stability\ indices\ obtained\ from\ the\ bootstrap\ samples}\label{tab:stability}\\

\ascline{1.5pt}{666666}{1-4}

\multicolumn{1}{>{\raggedright}m{\dimexpr 1.5in+0\tabcolsep}}{\textcolor[HTML]{000000}{\fontsize{10}{10}\selectfont{{Algorithm}}}} & \multicolumn{1}{>{\centering}m{\dimexpr 1.5in+0\tabcolsep}}{\textcolor[HTML]{000000}{\fontsize{10}{10}\selectfont{{Range\ of\ the\ optimal\ number\ of\ clusters\ found\ in\ the\ bootstrap\ samples}}}} & \multicolumn{1}{>{\centering}m{\dimexpr 1.5in+0\tabcolsep}}{\textcolor[HTML]{000000}{\fontsize{10}{10}\selectfont{{Rand\ Index\ [0,\ 1]}}}} & \multicolumn{1}{>{\centering}m{\dimexpr 1.5in+0\tabcolsep}}{\textcolor[HTML]{000000}{\fontsize{10}{10}\selectfont{{Adjusted\ Rand\ Index\ [-1,\ 1]}}}} \\

\ascline{1.5pt}{666666}{1-4}\endhead

\multicolumn{1}{>{\raggedright}m{\dimexpr 1.5in+0\tabcolsep}}{\textcolor[HTML]{000000}{\fontsize{10}{10}\selectfont{{AG-Eucl-Sil}}}} & \multicolumn{1}{>{\centering}m{\dimexpr 1.5in+0\tabcolsep}}{\textcolor[HTML]{000000}{\fontsize{10}{10}\selectfont{{2-3}}}} & \multicolumn{1}{>{\centering}m{\dimexpr 1.5in+0\tabcolsep}}{\textcolor[HTML]{000000}{\fontsize{10}{10}\selectfont{{0.96}}}} & \multicolumn{1}{>{\centering}m{\dimexpr 1.5in+0\tabcolsep}}{\textcolor[HTML]{000000}{\fontsize{10}{10}\selectfont{{0.88}}}} \\

\multicolumn{1}{>{\raggedright}m{\dimexpr 1.5in+0\tabcolsep}}{\textcolor[HTML]{000000}{\fontsize{10}{10}\selectfont{{AG-Eucl-Gap}}}} & \multicolumn{1}{>{\centering}m{\dimexpr 1.5in+0\tabcolsep}}{\textcolor[HTML]{000000}{\fontsize{10}{10}\selectfont{{1-2}}}} & \multicolumn{1}{>{\centering}m{\dimexpr 1.5in+0\tabcolsep}}{\textcolor[HTML]{000000}{\fontsize{10}{10}\selectfont{{0.90}}}} & \multicolumn{1}{>{\centering}m{\dimexpr 1.5in+0\tabcolsep}}{\textcolor[HTML]{000000}{\fontsize{10}{10}\selectfont{{0.61}}}} \\

\multicolumn{1}{>{\raggedright}m{\dimexpr 1.5in+0\tabcolsep}}{\textcolor[HTML]{000000}{\fontsize{10}{10}\selectfont{{AG-Gow-Sil}}}} & \multicolumn{1}{>{\centering}m{\dimexpr 1.5in+0\tabcolsep}}{\textcolor[HTML]{000000}{\fontsize{10}{10}\selectfont{{2}}}} & \multicolumn{1}{>{\centering}m{\dimexpr 1.5in+0\tabcolsep}}{\textcolor[HTML]{000000}{\fontsize{10}{10}\selectfont{{0.95}}}} & \multicolumn{1}{>{\centering}m{\dimexpr 1.5in+0\tabcolsep}}{\textcolor[HTML]{000000}{\fontsize{10}{10}\selectfont{{0.69}}}} \\

\multicolumn{1}{>{\raggedright}m{\dimexpr 1.5in+0\tabcolsep}}{\textcolor[HTML]{000000}{\fontsize{10}{10}\selectfont{{AG-Gow-Gap}}}} & \multicolumn{1}{>{\centering}m{\dimexpr 1.5in+0\tabcolsep}}{\textcolor[HTML]{000000}{\fontsize{10}{10}\selectfont{{1-2}}}} & \multicolumn{1}{>{\centering}m{\dimexpr 1.5in+0\tabcolsep}}{\textcolor[HTML]{000000}{\fontsize{10}{10}\selectfont{{0.99}}}} & \multicolumn{1}{>{\centering}m{\dimexpr 1.5in+0\tabcolsep}}{\textcolor[HTML]{000000}{\fontsize{10}{10}\selectfont{{0.88}}}} \\

\multicolumn{1}{>{\raggedright}m{\dimexpr 1.5in+0\tabcolsep}}{\textcolor[HTML]{000000}{\fontsize{10}{10}\selectfont{{KM-Eucl-Sil}}}} & \multicolumn{1}{>{\centering}m{\dimexpr 1.5in+0\tabcolsep}}{\textcolor[HTML]{000000}{\fontsize{10}{10}\selectfont{{2-10}}}} & \multicolumn{1}{>{\centering}m{\dimexpr 1.5in+0\tabcolsep}}{\textcolor[HTML]{000000}{\fontsize{10}{10}\selectfont{{0.80}}}} & \multicolumn{1}{>{\centering}m{\dimexpr 1.5in+0\tabcolsep}}{\textcolor[HTML]{000000}{\fontsize{10}{10}\selectfont{{0.51}}}} \\

\multicolumn{1}{>{\raggedright}m{\dimexpr 1.5in+0\tabcolsep}}{\textcolor[HTML]{000000}{\fontsize{10}{10}\selectfont{{KM-Eucl-Gap}}}} & \multicolumn{1}{>{\centering}m{\dimexpr 1.5in+0\tabcolsep}}{\textcolor[HTML]{000000}{\fontsize{10}{10}\selectfont{{1}}}} & \multicolumn{1}{>{\centering}m{\dimexpr 1.5in+0\tabcolsep}}{\textcolor[HTML]{000000}{\fontsize{10}{10}\selectfont{{1.00}}}} & \multicolumn{1}{>{\centering}m{\dimexpr 1.5in+0\tabcolsep}}{\textcolor[HTML]{000000}{\fontsize{10}{10}\selectfont{{1.00}}}} \\

\multicolumn{1}{>{\raggedright}m{\dimexpr 1.5in+0\tabcolsep}}{\textcolor[HTML]{000000}{\fontsize{10}{10}\selectfont{{KM-Gow-Sil}}}} & \multicolumn{1}{>{\centering}m{\dimexpr 1.5in+0\tabcolsep}}{\textcolor[HTML]{000000}{\fontsize{10}{10}\selectfont{{2-18}}}} & \multicolumn{1}{>{\centering}m{\dimexpr 1.5in+0\tabcolsep}}{\textcolor[HTML]{000000}{\fontsize{10}{10}\selectfont{{0.80}}}} & \multicolumn{1}{>{\centering}m{\dimexpr 1.5in+0\tabcolsep}}{\textcolor[HTML]{000000}{\fontsize{10}{10}\selectfont{{0.55}}}} \\

\multicolumn{1}{>{\raggedright}m{\dimexpr 1.5in+0\tabcolsep}}{\textcolor[HTML]{000000}{\fontsize{10}{10}\selectfont{{KM-Gow-Gap}}}} & \multicolumn{1}{>{\centering}m{\dimexpr 1.5in+0\tabcolsep}}{\textcolor[HTML]{000000}{\fontsize{10}{10}\selectfont{{1-3}}}} & \multicolumn{1}{>{\centering}m{\dimexpr 1.5in+0\tabcolsep}}{\textcolor[HTML]{000000}{\fontsize{10}{10}\selectfont{{0.42}}}} & \multicolumn{1}{>{\centering}m{\dimexpr 1.5in+0\tabcolsep}}{\textcolor[HTML]{000000}{\fontsize{10}{10}\selectfont{{0.13}}}} \\

\multicolumn{1}{>{\raggedright}m{\dimexpr 1.5in+0\tabcolsep}}{\textcolor[HTML]{000000}{\fontsize{10}{10}\selectfont{{SP-Eucl-Sil}}}} & \multicolumn{1}{>{\centering}m{\dimexpr 1.5in+0\tabcolsep}}{\textcolor[HTML]{000000}{\fontsize{10}{10}\selectfont{{2-25}}}} & \multicolumn{1}{>{\centering}m{\dimexpr 1.5in+0\tabcolsep}}{\textcolor[HTML]{000000}{\fontsize{10}{10}\selectfont{{0.78}}}} & \multicolumn{1}{>{\centering}m{\dimexpr 1.5in+0\tabcolsep}}{\textcolor[HTML]{000000}{\fontsize{10}{10}\selectfont{{0.49}}}} \\

\multicolumn{1}{>{\raggedright}m{\dimexpr 1.5in+0\tabcolsep}}{\textcolor[HTML]{000000}{\fontsize{10}{10}\selectfont{{SP-Eucl-Gap}}}} & \multicolumn{1}{>{\centering}m{\dimexpr 1.5in+0\tabcolsep}}{\textcolor[HTML]{000000}{\fontsize{10}{10}\selectfont{{1}}}} & \multicolumn{1}{>{\centering}m{\dimexpr 1.5in+0\tabcolsep}}{\textcolor[HTML]{000000}{\fontsize{10}{10}\selectfont{{1.00}}}} & \multicolumn{1}{>{\centering}m{\dimexpr 1.5in+0\tabcolsep}}{\textcolor[HTML]{000000}{\fontsize{10}{10}\selectfont{{1.00}}}} \\

\multicolumn{1}{>{\raggedright}m{\dimexpr 1.5in+0\tabcolsep}}{\textcolor[HTML]{000000}{\fontsize{10}{10}\selectfont{{SP-Gow-Sil}}}} & \multicolumn{1}{>{\centering}m{\dimexpr 1.5in+0\tabcolsep}}{\textcolor[HTML]{000000}{\fontsize{10}{10}\selectfont{{2}}}} & \multicolumn{1}{>{\centering}m{\dimexpr 1.5in+0\tabcolsep}}{\textcolor[HTML]{000000}{\fontsize{10}{10}\selectfont{{0.98}}}} & \multicolumn{1}{>{\centering}m{\dimexpr 1.5in+0\tabcolsep}}{\textcolor[HTML]{000000}{\fontsize{10}{10}\selectfont{{0.95}}}} \\

\multicolumn{1}{>{\raggedright}m{\dimexpr 1.5in+0\tabcolsep}}{\textcolor[HTML]{000000}{\fontsize{10}{10}\selectfont{{SP-Gow-Gap}}}} & \multicolumn{1}{>{\centering}m{\dimexpr 1.5in+0\tabcolsep}}{\textcolor[HTML]{000000}{\fontsize{10}{10}\selectfont{{6-8}}}} & \multicolumn{1}{>{\centering}m{\dimexpr 1.5in+0\tabcolsep}}{\textcolor[HTML]{000000}{\fontsize{10}{10}\selectfont{{0.91}}}} & \multicolumn{1}{>{\centering}m{\dimexpr 1.5in+0\tabcolsep}}{\textcolor[HTML]{000000}{\fontsize{10}{10}\selectfont{{0.67}}}} \\

\ascline{1.5pt}{666666}{1-4}

\end{longtable}

\arrayrulecolor[HTML]{000000}

\global\setlength{\arrayrulewidth}{\Oldarrayrulewidth}

\global\setlength{\tabcolsep}{\Oldtabcolsep}

\renewcommand*{\arraystretch}{1}

\clearpage

\section*{Figures}\label{figures}
\addcontentsline{toc}{section}{Figures}

\newpage

\renewcommand{\tablename}{Supplementary Table}
\setcounter{table}{0}

\renewcommand{\figurename}{Supplementary Figure}
\setcounter{figure}{0}

\section*{Supplementary Information}\label{supplementary-information}
\addcontentsline{toc}{section}{Supplementary Information}

\section*{Supplementary Tables}\label{supplementary-tables}
\addcontentsline{toc}{section}{Supplementary Tables}

\global\setlength{\Oldarrayrulewidth}{\arrayrulewidth}

\global\setlength{\Oldtabcolsep}{\tabcolsep}

\setlength{\tabcolsep}{2pt}

\renewcommand*{\arraystretch}{1.5}

\providecommand{\ascline}[3]{\noalign{\global\arrayrulewidth #1}\arrayrulecolor[HTML]{#2}\cline{#3}}

\begin{longtable}[c]{|p{1.25in}|p{1.25in}|p{1.25in}|p{1.25in}|p{1.25in}}

\caption{Results\ for\ a\ single\ run\ on\ the\ original\ dataset\ for\ each\ clustering\ algorithm}\label{tab:res}\\

\ascline{1.5pt}{666666}{1-5}

\multicolumn{1}{>{\raggedright}m{\dimexpr 1.25in+0\tabcolsep}}{\textcolor[HTML]{000000}{\fontsize{10}{10}\selectfont{{Algorithm}}}} & \multicolumn{1}{>{\centering}m{\dimexpr 1.25in+0\tabcolsep}}{\textcolor[HTML]{000000}{\fontsize{10}{10}\selectfont{{Optimal\ number\ of\ clusters}}}} & \multicolumn{1}{>{\centering}m{\dimexpr 1.25in+0\tabcolsep}}{\textcolor[HTML]{000000}{\fontsize{10}{10}\selectfont{{Smallest\ cluster\ size}}}} & \multicolumn{1}{>{\centering}m{\dimexpr 1.25in+0\tabcolsep}}{\textcolor[HTML]{000000}{\fontsize{10}{10}\selectfont{{Largest\ cluster\ size}}}} & \multicolumn{1}{>{\centering}m{\dimexpr 1.25in+0\tabcolsep}}{\textcolor[HTML]{000000}{\fontsize{10}{10}\selectfont{{Computation\ time\ (in\ minutes)}}}} \\

\ascline{1.5pt}{666666}{1-5}\endfirsthead \caption[]{Results\ for\ a\ single\ run\ on\ the\ original\ dataset\ for\ each\ clustering\ algorithm}\label{tab:res}\\

\ascline{1.5pt}{666666}{1-5}

\multicolumn{1}{>{\raggedright}m{\dimexpr 1.25in+0\tabcolsep}}{\textcolor[HTML]{000000}{\fontsize{10}{10}\selectfont{{Algorithm}}}} & \multicolumn{1}{>{\centering}m{\dimexpr 1.25in+0\tabcolsep}}{\textcolor[HTML]{000000}{\fontsize{10}{10}\selectfont{{Optimal\ number\ of\ clusters}}}} & \multicolumn{1}{>{\centering}m{\dimexpr 1.25in+0\tabcolsep}}{\textcolor[HTML]{000000}{\fontsize{10}{10}\selectfont{{Smallest\ cluster\ size}}}} & \multicolumn{1}{>{\centering}m{\dimexpr 1.25in+0\tabcolsep}}{\textcolor[HTML]{000000}{\fontsize{10}{10}\selectfont{{Largest\ cluster\ size}}}} & \multicolumn{1}{>{\centering}m{\dimexpr 1.25in+0\tabcolsep}}{\textcolor[HTML]{000000}{\fontsize{10}{10}\selectfont{{Computation\ time\ (in\ minutes)}}}} \\

\ascline{1.5pt}{666666}{1-5}\endhead

\multicolumn{1}{>{\raggedright}m{\dimexpr 1.25in+0\tabcolsep}}{\textcolor[HTML]{000000}{\fontsize{10}{10}\selectfont{{AG-Eucl-Gap}}}} & \multicolumn{1}{>{\centering}m{\dimexpr 1.25in+0\tabcolsep}}{\textcolor[HTML]{000000}{\fontsize{10}{10}\selectfont{{1}}}} & \multicolumn{1}{>{\centering}m{\dimexpr 1.25in+0\tabcolsep}}{\textcolor[HTML]{000000}{\fontsize{10}{10}\selectfont{{4,509}}}} & \multicolumn{1}{>{\centering}m{\dimexpr 1.25in+0\tabcolsep}}{\textcolor[HTML]{000000}{\fontsize{10}{10}\selectfont{{4,509}}}} & \multicolumn{1}{>{\centering}m{\dimexpr 1.25in+0\tabcolsep}}{\textcolor[HTML]{000000}{\fontsize{10}{10}\selectfont{{15.99}}}} \\

\multicolumn{1}{>{\raggedright}m{\dimexpr 1.25in+0\tabcolsep}}{\textcolor[HTML]{000000}{\fontsize{10}{10}\selectfont{{AG-Eucl-Sil}}}} & \multicolumn{1}{>{\centering}m{\dimexpr 1.25in+0\tabcolsep}}{\textcolor[HTML]{000000}{\fontsize{10}{10}\selectfont{{3}}}} & \multicolumn{1}{>{\centering}m{\dimexpr 1.25in+0\tabcolsep}}{\textcolor[HTML]{000000}{\fontsize{10}{10}\selectfont{{20}}}} & \multicolumn{1}{>{\centering}m{\dimexpr 1.25in+0\tabcolsep}}{\textcolor[HTML]{000000}{\fontsize{10}{10}\selectfont{{3,679}}}} & \multicolumn{1}{>{\centering}m{\dimexpr 1.25in+0\tabcolsep}}{\textcolor[HTML]{000000}{\fontsize{10}{10}\selectfont{{0.12}}}} \\

\multicolumn{1}{>{\raggedright}m{\dimexpr 1.25in+0\tabcolsep}}{\textcolor[HTML]{000000}{\fontsize{10}{10}\selectfont{{AG-Gow-Gap}}}} & \multicolumn{1}{>{\centering}m{\dimexpr 1.25in+0\tabcolsep}}{\textcolor[HTML]{000000}{\fontsize{10}{10}\selectfont{{1}}}} & \multicolumn{1}{>{\centering}m{\dimexpr 1.25in+0\tabcolsep}}{\textcolor[HTML]{000000}{\fontsize{10}{10}\selectfont{{4,509}}}} & \multicolumn{1}{>{\centering}m{\dimexpr 1.25in+0\tabcolsep}}{\textcolor[HTML]{000000}{\fontsize{10}{10}\selectfont{{4,509}}}} & \multicolumn{1}{>{\centering}m{\dimexpr 1.25in+0\tabcolsep}}{\textcolor[HTML]{000000}{\fontsize{10}{10}\selectfont{{22.85}}}} \\

\multicolumn{1}{>{\raggedright}m{\dimexpr 1.25in+0\tabcolsep}}{\textcolor[HTML]{000000}{\fontsize{10}{10}\selectfont{{AG-Gow-Sil}}}} & \multicolumn{1}{>{\centering}m{\dimexpr 1.25in+0\tabcolsep}}{\textcolor[HTML]{000000}{\fontsize{10}{10}\selectfont{{2}}}} & \multicolumn{1}{>{\centering}m{\dimexpr 1.25in+0\tabcolsep}}{\textcolor[HTML]{000000}{\fontsize{10}{10}\selectfont{{202}}}} & \multicolumn{1}{>{\centering}m{\dimexpr 1.25in+0\tabcolsep}}{\textcolor[HTML]{000000}{\fontsize{10}{10}\selectfont{{4,307}}}} & \multicolumn{1}{>{\centering}m{\dimexpr 1.25in+0\tabcolsep}}{\textcolor[HTML]{000000}{\fontsize{10}{10}\selectfont{{0.31}}}} \\

\multicolumn{1}{>{\raggedright}m{\dimexpr 1.25in+0\tabcolsep}}{\textcolor[HTML]{000000}{\fontsize{10}{10}\selectfont{{KM-Eucl-Gap}}}} & \multicolumn{1}{>{\centering}m{\dimexpr 1.25in+0\tabcolsep}}{\textcolor[HTML]{000000}{\fontsize{10}{10}\selectfont{{1}}}} & \multicolumn{1}{>{\centering}m{\dimexpr 1.25in+0\tabcolsep}}{\textcolor[HTML]{000000}{\fontsize{10}{10}\selectfont{{4,509}}}} & \multicolumn{1}{>{\centering}m{\dimexpr 1.25in+0\tabcolsep}}{\textcolor[HTML]{000000}{\fontsize{10}{10}\selectfont{{4,509}}}} & \multicolumn{1}{>{\centering}m{\dimexpr 1.25in+0\tabcolsep}}{\textcolor[HTML]{000000}{\fontsize{10}{10}\selectfont{{10.10}}}} \\

\multicolumn{1}{>{\raggedright}m{\dimexpr 1.25in+0\tabcolsep}}{\textcolor[HTML]{000000}{\fontsize{10}{10}\selectfont{{KM-Eucl-Sil}}}} & \multicolumn{1}{>{\centering}m{\dimexpr 1.25in+0\tabcolsep}}{\textcolor[HTML]{000000}{\fontsize{10}{10}\selectfont{{8}}}} & \multicolumn{1}{>{\centering}m{\dimexpr 1.25in+0\tabcolsep}}{\textcolor[HTML]{000000}{\fontsize{10}{10}\selectfont{{263}}}} & \multicolumn{1}{>{\centering}m{\dimexpr 1.25in+0\tabcolsep}}{\textcolor[HTML]{000000}{\fontsize{10}{10}\selectfont{{869}}}} & \multicolumn{1}{>{\centering}m{\dimexpr 1.25in+0\tabcolsep}}{\textcolor[HTML]{000000}{\fontsize{10}{10}\selectfont{{0.11}}}} \\

\multicolumn{1}{>{\raggedright}m{\dimexpr 1.25in+0\tabcolsep}}{\textcolor[HTML]{000000}{\fontsize{10}{10}\selectfont{{KM-Gow-Gap}}}} & \multicolumn{1}{>{\centering}m{\dimexpr 1.25in+0\tabcolsep}}{\textcolor[HTML]{000000}{\fontsize{10}{10}\selectfont{{6}}}} & \multicolumn{1}{>{\centering}m{\dimexpr 1.25in+0\tabcolsep}}{\textcolor[HTML]{000000}{\fontsize{10}{10}\selectfont{{460}}}} & \multicolumn{1}{>{\centering}m{\dimexpr 1.25in+0\tabcolsep}}{\textcolor[HTML]{000000}{\fontsize{10}{10}\selectfont{{1,057}}}} & \multicolumn{1}{>{\centering}m{\dimexpr 1.25in+0\tabcolsep}}{\textcolor[HTML]{000000}{\fontsize{10}{10}\selectfont{{6.08}}}} \\

\multicolumn{1}{>{\raggedright}m{\dimexpr 1.25in+0\tabcolsep}}{\textcolor[HTML]{000000}{\fontsize{10}{10}\selectfont{{KM-Gow-Sil}}}} & \multicolumn{1}{>{\centering}m{\dimexpr 1.25in+0\tabcolsep}}{\textcolor[HTML]{000000}{\fontsize{10}{10}\selectfont{{4}}}} & \multicolumn{1}{>{\centering}m{\dimexpr 1.25in+0\tabcolsep}}{\textcolor[HTML]{000000}{\fontsize{10}{10}\selectfont{{773}}}} & \multicolumn{1}{>{\centering}m{\dimexpr 1.25in+0\tabcolsep}}{\textcolor[HTML]{000000}{\fontsize{10}{10}\selectfont{{1,615}}}} & \multicolumn{1}{>{\centering}m{\dimexpr 1.25in+0\tabcolsep}}{\textcolor[HTML]{000000}{\fontsize{10}{10}\selectfont{{0.25}}}} \\

\multicolumn{1}{>{\raggedright}m{\dimexpr 1.25in+0\tabcolsep}}{\textcolor[HTML]{000000}{\fontsize{10}{10}\selectfont{{SP-Eucl-Gap}}}} & \multicolumn{1}{>{\centering}m{\dimexpr 1.25in+0\tabcolsep}}{\textcolor[HTML]{000000}{\fontsize{10}{10}\selectfont{{1}}}} & \multicolumn{1}{>{\centering}m{\dimexpr 1.25in+0\tabcolsep}}{\textcolor[HTML]{000000}{\fontsize{10}{10}\selectfont{{4,509}}}} & \multicolumn{1}{>{\centering}m{\dimexpr 1.25in+0\tabcolsep}}{\textcolor[HTML]{000000}{\fontsize{10}{10}\selectfont{{4,509}}}} & \multicolumn{1}{>{\centering}m{\dimexpr 1.25in+0\tabcolsep}}{\textcolor[HTML]{000000}{\fontsize{10}{10}\selectfont{{135.56}}}} \\

\multicolumn{1}{>{\raggedright}m{\dimexpr 1.25in+0\tabcolsep}}{\textcolor[HTML]{000000}{\fontsize{10}{10}\selectfont{{SP-Eucl-Sil}}}} & \multicolumn{1}{>{\centering}m{\dimexpr 1.25in+0\tabcolsep}}{\textcolor[HTML]{000000}{\fontsize{10}{10}\selectfont{{25}}}} & \multicolumn{1}{>{\centering}m{\dimexpr 1.25in+0\tabcolsep}}{\textcolor[HTML]{000000}{\fontsize{10}{10}\selectfont{{10}}}} & \multicolumn{1}{>{\centering}m{\dimexpr 1.25in+0\tabcolsep}}{\textcolor[HTML]{000000}{\fontsize{10}{10}\selectfont{{1,951}}}} & \multicolumn{1}{>{\centering}m{\dimexpr 1.25in+0\tabcolsep}}{\textcolor[HTML]{000000}{\fontsize{10}{10}\selectfont{{0.23}}}} \\

\multicolumn{1}{>{\raggedright}m{\dimexpr 1.25in+0\tabcolsep}}{\textcolor[HTML]{000000}{\fontsize{10}{10}\selectfont{{SP-Gow-Gap}}}} & \multicolumn{1}{>{\centering}m{\dimexpr 1.25in+0\tabcolsep}}{\textcolor[HTML]{000000}{\fontsize{10}{10}\selectfont{{7}}}} & \multicolumn{1}{>{\centering}m{\dimexpr 1.25in+0\tabcolsep}}{\textcolor[HTML]{000000}{\fontsize{10}{10}\selectfont{{143}}}} & \multicolumn{1}{>{\centering}m{\dimexpr 1.25in+0\tabcolsep}}{\textcolor[HTML]{000000}{\fontsize{10}{10}\selectfont{{1,076}}}} & \multicolumn{1}{>{\centering}m{\dimexpr 1.25in+0\tabcolsep}}{\textcolor[HTML]{000000}{\fontsize{10}{10}\selectfont{{118.47}}}} \\

\multicolumn{1}{>{\raggedright}m{\dimexpr 1.25in+0\tabcolsep}}{\textcolor[HTML]{000000}{\fontsize{10}{10}\selectfont{{SP-Gow-Sil}}}} & \multicolumn{1}{>{\centering}m{\dimexpr 1.25in+0\tabcolsep}}{\textcolor[HTML]{000000}{\fontsize{10}{10}\selectfont{{2}}}} & \multicolumn{1}{>{\centering}m{\dimexpr 1.25in+0\tabcolsep}}{\textcolor[HTML]{000000}{\fontsize{10}{10}\selectfont{{839}}}} & \multicolumn{1}{>{\centering}m{\dimexpr 1.25in+0\tabcolsep}}{\textcolor[HTML]{000000}{\fontsize{10}{10}\selectfont{{3,670}}}} & \multicolumn{1}{>{\centering}m{\dimexpr 1.25in+0\tabcolsep}}{\textcolor[HTML]{000000}{\fontsize{10}{10}\selectfont{{0.67}}}} \\

\ascline{1.5pt}{666666}{1-5}

\end{longtable}

\arrayrulecolor[HTML]{000000}

\global\setlength{\arrayrulewidth}{\Oldarrayrulewidth}

\global\setlength{\tabcolsep}{\Oldtabcolsep}

\renewcommand*{\arraystretch}{1}

\footnotesize

Computation time was measured on a MacBook Air with Apple M3 chip and 8GB RAM. KM was run using an alternate initialization method; AG used average linkage; SP used 3 degrees of freedom as hyperparameter.

\global\setlength{\Oldarrayrulewidth}{\arrayrulewidth}

\global\setlength{\Oldtabcolsep}{\tabcolsep}

\setlength{\tabcolsep}{2pt}

\renewcommand*{\arraystretch}{1.5}

\providecommand{\ascline}[3]{\noalign{\global\arrayrulewidth #1}\arrayrulecolor[HTML]{#2}\cline{#3}}

\begin{longtable}[c]{|p{1.50in}|p{1.25in}|p{1.25in}|p{2.50in}}

\caption{Bootstrap\ AUC\ results}\label{tab:bootresults}\\

\ascline{1.5pt}{666666}{1-4}

\multicolumn{1}{>{\raggedright}m{\dimexpr 1.5in+0\tabcolsep}}{\textcolor[HTML]{000000}{\fontsize{10}{10}\selectfont{{Algorithm}}}} & \multicolumn{1}{>{\centering}m{\dimexpr 1.25in+0\tabcolsep}}{\textcolor[HTML]{000000}{\fontsize{10}{10}\selectfont{{Apparent\ AUC}}}} & \multicolumn{1}{>{\centering}m{\dimexpr 1.25in+0\tabcolsep}}{\textcolor[HTML]{000000}{\fontsize{10}{10}\selectfont{{Optimism}}}} & \multicolumn{1}{>{\centering}m{\dimexpr 2.5in+0\tabcolsep}}{\textcolor[HTML]{000000}{\fontsize{10}{10}\selectfont{{Optimism-adjusted\ AUC\ (95\%\ CI)}}}} \\

\ascline{1.5pt}{666666}{1-4}\endfirsthead \caption[]{Bootstrap\ AUC\ results}\label{tab:bootresults}\\

\ascline{1.5pt}{666666}{1-4}

\multicolumn{1}{>{\raggedright}m{\dimexpr 1.5in+0\tabcolsep}}{\textcolor[HTML]{000000}{\fontsize{10}{10}\selectfont{{Algorithm}}}} & \multicolumn{1}{>{\centering}m{\dimexpr 1.25in+0\tabcolsep}}{\textcolor[HTML]{000000}{\fontsize{10}{10}\selectfont{{Apparent\ AUC}}}} & \multicolumn{1}{>{\centering}m{\dimexpr 1.25in+0\tabcolsep}}{\textcolor[HTML]{000000}{\fontsize{10}{10}\selectfont{{Optimism}}}} & \multicolumn{1}{>{\centering}m{\dimexpr 2.5in+0\tabcolsep}}{\textcolor[HTML]{000000}{\fontsize{10}{10}\selectfont{{Optimism-adjusted\ AUC\ (95\%\ CI)}}}} \\

\ascline{1.5pt}{666666}{1-4}\endhead

\multicolumn{1}{>{\raggedright}m{\dimexpr 1.5in+0\tabcolsep}}{\textcolor[HTML]{000000}{\fontsize{10}{10}\selectfont{{AG-Eucl-Gap}}}} & \multicolumn{1}{>{\centering}m{\dimexpr 1.25in+0\tabcolsep}}{\textcolor[HTML]{000000}{\fontsize{10}{10}\selectfont{{0.500}}}} & \multicolumn{1}{>{\centering}m{\dimexpr 1.25in+0\tabcolsep}}{\textcolor[HTML]{000000}{\fontsize{10}{10}\selectfont{{0.063}}}} & \multicolumn{1}{>{\centering}m{\dimexpr 2.5in+0\tabcolsep}}{\textcolor[HTML]{000000}{\fontsize{10}{10}\selectfont{{0.437\ (0.258,\ 0.616)}}}} \\

\multicolumn{1}{>{\raggedright}m{\dimexpr 1.5in+0\tabcolsep}}{\textcolor[HTML]{000000}{\fontsize{10}{10}\selectfont{{AG-Eucl-Sil}}}} & \multicolumn{1}{>{\centering}m{\dimexpr 1.25in+0\tabcolsep}}{\textcolor[HTML]{000000}{\fontsize{10}{10}\selectfont{{0.697}}}} & \multicolumn{1}{>{\centering}m{\dimexpr 1.25in+0\tabcolsep}}{\textcolor[HTML]{000000}{\fontsize{10}{10}\selectfont{{0.171}}}} & \multicolumn{1}{>{\centering}m{\dimexpr 2.5in+0\tabcolsep}}{\textcolor[HTML]{000000}{\fontsize{10}{10}\selectfont{{0.527\ (0.437,\ 0.616)}}}} \\

\multicolumn{1}{>{\raggedright}m{\dimexpr 1.5in+0\tabcolsep}}{\textcolor[HTML]{000000}{\fontsize{10}{10}\selectfont{{AG-Gow-Gap}}}} & \multicolumn{1}{>{\centering}m{\dimexpr 1.25in+0\tabcolsep}}{\textcolor[HTML]{000000}{\fontsize{10}{10}\selectfont{{0.500}}}} & \multicolumn{1}{>{\centering}m{\dimexpr 1.25in+0\tabcolsep}}{\textcolor[HTML]{000000}{\fontsize{10}{10}\selectfont{{-0.002}}}} & \multicolumn{1}{>{\centering}m{\dimexpr 2.5in+0\tabcolsep}}{\textcolor[HTML]{000000}{\fontsize{10}{10}\selectfont{{0.502\ (0.48,\ 0.524)}}}} \\

\multicolumn{1}{>{\raggedright}m{\dimexpr 1.5in+0\tabcolsep}}{\textcolor[HTML]{000000}{\fontsize{10}{10}\selectfont{{AG-Gow-Sil}}}} & \multicolumn{1}{>{\centering}m{\dimexpr 1.25in+0\tabcolsep}}{\textcolor[HTML]{000000}{\fontsize{10}{10}\selectfont{{0.554}}}} & \multicolumn{1}{>{\centering}m{\dimexpr 1.25in+0\tabcolsep}}{\textcolor[HTML]{000000}{\fontsize{10}{10}\selectfont{{-0.003}}}} & \multicolumn{1}{>{\centering}m{\dimexpr 2.5in+0\tabcolsep}}{\textcolor[HTML]{000000}{\fontsize{10}{10}\selectfont{{0.557\ (0.529,\ 0.584)}}}} \\

\multicolumn{1}{>{\raggedright}m{\dimexpr 1.5in+0\tabcolsep}}{\textcolor[HTML]{000000}{\fontsize{10}{10}\selectfont{{KM-Eucl-Gap}}}} & \multicolumn{1}{>{\centering}m{\dimexpr 1.25in+0\tabcolsep}}{\textcolor[HTML]{000000}{\fontsize{10}{10}\selectfont{{0.500}}}} & \multicolumn{1}{>{\centering}m{\dimexpr 1.25in+0\tabcolsep}}{\textcolor[HTML]{000000}{\fontsize{10}{10}\selectfont{{0.000}}}} & \multicolumn{1}{>{\centering}m{\dimexpr 2.5in+0\tabcolsep}}{\textcolor[HTML]{000000}{\fontsize{10}{10}\selectfont{{0.5\ (0.5,\ 0.5)}}}} \\

\multicolumn{1}{>{\raggedright}m{\dimexpr 1.5in+0\tabcolsep}}{\textcolor[HTML]{000000}{\fontsize{10}{10}\selectfont{{KM-Eucl-Sil}}}} & \multicolumn{1}{>{\centering}m{\dimexpr 1.25in+0\tabcolsep}}{\textcolor[HTML]{000000}{\fontsize{10}{10}\selectfont{{0.776}}}} & \multicolumn{1}{>{\centering}m{\dimexpr 1.25in+0\tabcolsep}}{\textcolor[HTML]{000000}{\fontsize{10}{10}\selectfont{{0.103}}}} & \multicolumn{1}{>{\centering}m{\dimexpr 2.5in+0\tabcolsep}}{\textcolor[HTML]{000000}{\fontsize{10}{10}\selectfont{{0.673\ (0.582,\ 0.764)}}}} \\

\multicolumn{1}{>{\raggedright}m{\dimexpr 1.5in+0\tabcolsep}}{\textcolor[HTML]{000000}{\fontsize{10}{10}\selectfont{{KM-Gow-Gap}}}} & \multicolumn{1}{>{\centering}m{\dimexpr 1.25in+0\tabcolsep}}{\textcolor[HTML]{000000}{\fontsize{10}{10}\selectfont{{0.711}}}} & \multicolumn{1}{>{\centering}m{\dimexpr 1.25in+0\tabcolsep}}{\textcolor[HTML]{000000}{\fontsize{10}{10}\selectfont{{0.015}}}} & \multicolumn{1}{>{\centering}m{\dimexpr 2.5in+0\tabcolsep}}{\textcolor[HTML]{000000}{\fontsize{10}{10}\selectfont{{0.697\ (0.623,\ 0.771)}}}} \\

\multicolumn{1}{>{\raggedright}m{\dimexpr 1.5in+0\tabcolsep}}{\textcolor[HTML]{000000}{\fontsize{10}{10}\selectfont{{KM-Gow-Sil}}}} & \multicolumn{1}{>{\centering}m{\dimexpr 1.25in+0\tabcolsep}}{\textcolor[HTML]{000000}{\fontsize{10}{10}\selectfont{{0.648}}}} & \multicolumn{1}{>{\centering}m{\dimexpr 1.25in+0\tabcolsep}}{\textcolor[HTML]{000000}{\fontsize{10}{10}\selectfont{{0.080}}}} & \multicolumn{1}{>{\centering}m{\dimexpr 2.5in+0\tabcolsep}}{\textcolor[HTML]{000000}{\fontsize{10}{10}\selectfont{{0.567\ (0.346,\ 0.788)}}}} \\

\multicolumn{1}{>{\raggedright}m{\dimexpr 1.5in+0\tabcolsep}}{\textcolor[HTML]{000000}{\fontsize{10}{10}\selectfont{{SP-Eucl-Gap}}}} & \multicolumn{1}{>{\centering}m{\dimexpr 1.25in+0\tabcolsep}}{\textcolor[HTML]{000000}{\fontsize{10}{10}\selectfont{{0.500}}}} & \multicolumn{1}{>{\centering}m{\dimexpr 1.25in+0\tabcolsep}}{\textcolor[HTML]{000000}{\fontsize{10}{10}\selectfont{{0.000}}}} & \multicolumn{1}{>{\centering}m{\dimexpr 2.5in+0\tabcolsep}}{\textcolor[HTML]{000000}{\fontsize{10}{10}\selectfont{{0.5\ (0.5,\ 0.5)}}}} \\

\multicolumn{1}{>{\raggedright}m{\dimexpr 1.5in+0\tabcolsep}}{\textcolor[HTML]{000000}{\fontsize{10}{10}\selectfont{{SP-Eucl-Sil}}}} & \multicolumn{1}{>{\centering}m{\dimexpr 1.25in+0\tabcolsep}}{\textcolor[HTML]{000000}{\fontsize{10}{10}\selectfont{{0.788}}}} & \multicolumn{1}{>{\centering}m{\dimexpr 1.25in+0\tabcolsep}}{\textcolor[HTML]{000000}{\fontsize{10}{10}\selectfont{{0.110}}}} & \multicolumn{1}{>{\centering}m{\dimexpr 2.5in+0\tabcolsep}}{\textcolor[HTML]{000000}{\fontsize{10}{10}\selectfont{{0.678\ (0.486,\ 0.869)}}}} \\

\multicolumn{1}{>{\raggedright}m{\dimexpr 1.5in+0\tabcolsep}}{\textcolor[HTML]{000000}{\fontsize{10}{10}\selectfont{{SP-Gow-Gap}}}} & \multicolumn{1}{>{\centering}m{\dimexpr 1.25in+0\tabcolsep}}{\textcolor[HTML]{000000}{\fontsize{10}{10}\selectfont{{0.747}}}} & \multicolumn{1}{>{\centering}m{\dimexpr 1.25in+0\tabcolsep}}{\textcolor[HTML]{000000}{\fontsize{10}{10}\selectfont{{0.218}}}} & \multicolumn{1}{>{\centering}m{\dimexpr 2.5in+0\tabcolsep}}{\textcolor[HTML]{000000}{\fontsize{10}{10}\selectfont{{0.529\ (0.496,\ 0.561)}}}} \\

\multicolumn{1}{>{\raggedright}m{\dimexpr 1.5in+0\tabcolsep}}{\textcolor[HTML]{000000}{\fontsize{10}{10}\selectfont{{SP-Gow-Sil}}}} & \multicolumn{1}{>{\centering}m{\dimexpr 1.25in+0\tabcolsep}}{\textcolor[HTML]{000000}{\fontsize{10}{10}\selectfont{{0.666}}}} & \multicolumn{1}{>{\centering}m{\dimexpr 1.25in+0\tabcolsep}}{\textcolor[HTML]{000000}{\fontsize{10}{10}\selectfont{{0.002}}}} & \multicolumn{1}{>{\centering}m{\dimexpr 2.5in+0\tabcolsep}}{\textcolor[HTML]{000000}{\fontsize{10}{10}\selectfont{{0.664\ (0.644,\ 0.684)}}}} \\

\multicolumn{1}{>{\raggedright}m{\dimexpr 1.5in+0\tabcolsep}}{\textcolor[HTML]{000000}{\fontsize{10}{10}\selectfont{{Log-Reg}}}} & \multicolumn{1}{>{\centering}m{\dimexpr 1.25in+0\tabcolsep}}{\textcolor[HTML]{000000}{\fontsize{10}{10}\selectfont{{0.872}}}} & \multicolumn{1}{>{\centering}m{\dimexpr 1.25in+0\tabcolsep}}{\textcolor[HTML]{000000}{\fontsize{10}{10}\selectfont{{0.001}}}} & \multicolumn{1}{>{\centering}m{\dimexpr 2.5in+0\tabcolsep}}{\textcolor[HTML]{000000}{\fontsize{10}{10}\selectfont{{0.871\ (0.87,\ 0.873)}}}} \\

\multicolumn{1}{>{\raggedright}m{\dimexpr 1.5in+0\tabcolsep}}{\textcolor[HTML]{000000}{\fontsize{10}{10}\selectfont{{Log-Reg-GCS}}}} & \multicolumn{1}{>{\centering}m{\dimexpr 1.25in+0\tabcolsep}}{\textcolor[HTML]{000000}{\fontsize{10}{10}\selectfont{{0.808}}}} & \multicolumn{1}{>{\centering}m{\dimexpr 1.25in+0\tabcolsep}}{\textcolor[HTML]{000000}{\fontsize{10}{10}\selectfont{{0.001}}}} & \multicolumn{1}{>{\centering}m{\dimexpr 2.5in+0\tabcolsep}}{\textcolor[HTML]{000000}{\fontsize{10}{10}\selectfont{{0.807\ (0.806,\ 0.808)}}}} \\

\ascline{1.5pt}{666666}{1-4}

\end{longtable}

\arrayrulecolor[HTML]{000000}

\global\setlength{\arrayrulewidth}{\Oldarrayrulewidth}

\global\setlength{\tabcolsep}{\Oldtabcolsep}

\renewcommand*{\arraystretch}{1}

\begin{footnotesize} Abbreviations: AG = agglomerative clustering; KM = K-Medoids clustering; SP = spectral clustering; Eucl = Euclidean distance; Gow = Gower's distance; Sil = silhouette value; Gap = gap statistic; Log-Reg = logistic regression; GCS = Glasgow Coma Scale Score \end{footnotesize}

\newpage

\section*{Supplementary Figures}\label{supplementary-figures}
\addcontentsline{toc}{section}{Supplementary Figures}

\begin{figure}

{\centering \includegraphics[width=0.95\linewidth]{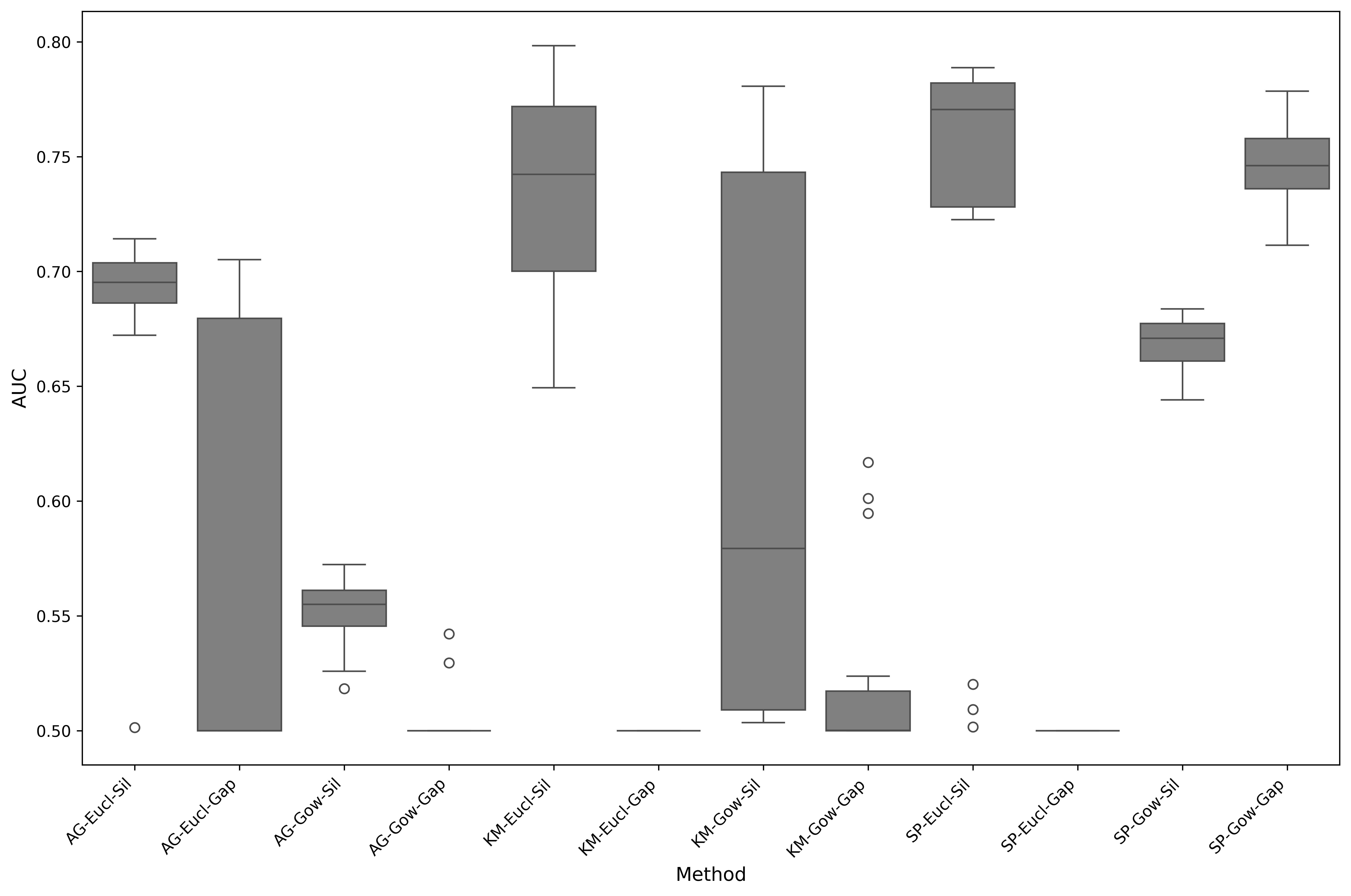} 

}

\caption{Bootstrap AUCs (i.e., discrimination of clustering solutions obtained in  bootstrap samples and applied to the original data), for 200 bootstrap samples}\label{fig:aucboot}
\end{figure}

\section*{Syntax}\label{syntax}
\addcontentsline{toc}{section}{Syntax}

The code used for the analyses is available on \href{https://github.com/aneetachacko/Clustering}{GitHub}.


\begin{CSLReferences}{0}{1}
\bibitem[\citeproctext]{ref-manley2025new}
\CSLLeftMargin{1. }%
\CSLRightInline{{Manley GT, Dams-O'Connor K, Alosco ML, et al.} A new characterisation of acute traumatic brain injury: The NIH-NINDS TBI classification and nomenclature initiative. \emph{The Lancet Neurology}. 2025;24(6):512-523.}

\bibitem[\citeproctext]{ref-tas2024unsupervised}
\CSLLeftMargin{2. }%
\CSLRightInline{Tas J, Rass V, Ianosi BA, Heidbreder A, Bergmann M, Helbok R. Unsupervised clustering in neurocritical care: A systematic review. \emph{Neurocritical Care}. Published online 2024:1-13.}

\bibitem[\citeproctext]{ref-gelman2013forkingpaths}
\CSLLeftMargin{3. }%
\CSLRightInline{Gelman A, Loken E. The garden of forking paths: Why multiple comparisons can be a problem, even when there is no {``fishing expedition''} or {``p-hacking''} and the research hypothesis was posited ahead of time. \emph{Department of Statistics, Columbia University}. 2013;348(1-17):3.}

\bibitem[\citeproctext]{ref-gravesteijn2020toward}
\CSLLeftMargin{4. }%
\CSLRightInline{Gravesteijn BY, Sewalt CA, Ercole A, et al. Toward a new multi-dimensional classification of traumatic brain injury: A collaborative european neurotrauma effectiveness research for traumatic brain injury study. \emph{Journal of Neurotrauma}. 2020;37(7):1002-1010.}

\bibitem[\citeproctext]{ref-pam1990}
\CSLLeftMargin{5. }%
\CSLRightInline{Kaufman L, Rousseeuw PJ. \emph{Finding Groups in Data: An Introduction to Cluster Analysis}. John Wiley \& Sons; 2009.}

\bibitem[\citeproctext]{ref-rousseeuw1987silhouettes}
\CSLLeftMargin{6. }%
\CSLRightInline{Rousseeuw PJ. Silhouettes: A graphical aid to the interpretation and validation of cluster analysis. \emph{Journal of computational and applied mathematics}. 1987;20:53-65.}

\bibitem[\citeproctext]{ref-gower1971similarity}
\CSLLeftMargin{7. }%
\CSLRightInline{Gower JC. A general coefficient of similarity and some of its properties. \emph{Biometrics}. Published online 1971:857-871.}

\bibitem[\citeproctext]{ref-maas2015center}
\CSLLeftMargin{8. }%
\CSLRightInline{{Maas AI, Menon DK, Steyerberg EW, et al.} Collaborative european neurotrauma effectiveness research in traumatic brain injury (CENTER-TBI): A prospective longitudinal observational study. \emph{Neurosurgery}. 2015;76(1):67-80.}

\bibitem[\citeproctext]{ref-steyerberg2019casemix}
\CSLLeftMargin{9. }%
\CSLRightInline{{Steyerberg EW, Wiegers E, Sewalt C, et al.} Case-mix, care pathways, and outcomes in patients with traumatic brain injury in CENTER-TBI: A european prospective, multicentre, longitudinal, cohort study. \emph{The Lancet Neurology}. 2019;18(10):923-934.}

\bibitem[\citeproctext]{ref-steyerberg2019heterogeneity}
\CSLLeftMargin{10. }%
\CSLRightInline{Steyerberg EW, Nieboer D, Debray TP, Houwelingen HC van. Assessment of heterogeneity in an individual participant data meta-analysis of prediction models: An overview and illustration. \emph{Statistics in Medicine}. 2019;38(22):4290-4309.}

\bibitem[\citeproctext]{ref-vanbuuren2018flexible}
\CSLLeftMargin{11. }%
\CSLRightInline{Van Buuren S. \emph{Flexible Imputation of Missing Data}. 2nd ed. CRC press Boca Raton, FL; 2012.}

\bibitem[\citeproctext]{ref-R-base}
\CSLLeftMargin{12. }%
\CSLRightInline{R Core Team. \emph{R: A Language and Environment for Statistical Computing}. R Foundation for Statistical Computing; 2024. \url{https://www.R-project.org/}}

\bibitem[\citeproctext]{ref-hastie2009elements}
\CSLLeftMargin{13. }%
\CSLRightInline{Hastie T, Tibshirani R, Friedman JH. \emph{The Elements of Statistical Learning: Data Mining, Inference, and Prediction}. 2nd ed. Springer; 2009.}

\bibitem[\citeproctext]{ref-scikit-learn}
\CSLLeftMargin{14. }%
\CSLRightInline{Pedregosa F, Varoquaux G, Gramfort A, et al. Scikit-learn: Machine learning in {P}ython. \emph{Journal of Machine Learning Research}. 2011;12:2825-2830.}

\bibitem[\citeproctext]{ref-jain1999data}
\CSLLeftMargin{15. }%
\CSLRightInline{Jain AK, Murty MN, Flynn PJ. Data clustering: A review. \emph{ACM computing surveys (CSUR)}. 1999;31(3):264-323.}

\bibitem[\citeproctext]{ref-xu2015comprehensive}
\CSLLeftMargin{16. }%
\CSLRightInline{Xu D, Tian Y. A comprehensive survey of clustering algorithms. \emph{Annals of data science}. 2015;2(2):165-193.}

\bibitem[\citeproctext]{ref-deza2009encyclopedia}
\CSLLeftMargin{17. }%
\CSLRightInline{Deza MM, Deza E. Encyclopedia of distances. In: \emph{Encyclopedia of Distances}. Springer; 2009:1-583.}

\bibitem[\citeproctext]{ref-tibshirani2001gap}
\CSLLeftMargin{18. }%
\CSLRightInline{Tibshirani R, Walther G, Hastie T. Estimating the number of clusters in a data set via the gap statistic. \emph{Journal of the royal statistical society: series B (statistical methodology)}. 2001;63(2):411-423.}

\bibitem[\citeproctext]{ref-von2012clustering}
\CSLLeftMargin{19. }%
\CSLRightInline{Von Luxburg U, Williamson RC, Guyon I. Clustering: Science or art? In: \emph{Proceedings of ICML Workshop on Unsupervised and Transfer Learning}. JMLR Workshop; Conference Proceedings; 2012:65-79.}

\bibitem[\citeproctext]{ref-liu2022stability}
\CSLLeftMargin{20. }%
\CSLRightInline{Liu T, Yu H, Blair RH. Stability estimation for unsupervised clustering: A review. \emph{Wiley Interdisciplinary Reviews: Computational Statistics}. 2022;14(6):e1575.}

\bibitem[\citeproctext]{ref-hubert1985comparing}
\CSLLeftMargin{21. }%
\CSLRightInline{Hubert L, Arabie P. Comparing partitions. \emph{Journal of classification}. 1985;2:193-218.}

\bibitem[\citeproctext]{ref-lex2014upset}
\CSLLeftMargin{22. }%
\CSLRightInline{Lex A, Gehlenborg N, Strobelt H, Vuillemot R, Pfister H. UpSet: Visualization of intersecting sets. \emph{IEEE transactions on visualization and computer graphics}. 2014;20(12):1983-1992.}

\bibitem[\citeproctext]{ref-steyerberg2001internal}
\CSLLeftMargin{23. }%
\CSLRightInline{Steyerberg EW, Harrell Jr FE, Borsboom GJ, Eijkemans M, Vergouwe Y, Habbema JDF. Internal validation of predictive models: Efficiency of some procedures for logistic regression analysis. \emph{Journal of clinical epidemiology}. 2001;54(8):774-781.}

\bibitem[\citeproctext]{ref-van2016modern}
\CSLLeftMargin{24. }%
\CSLRightInline{Ploeg T van der, Nieboer D, Steyerberg EW. Modern modeling techniques had limited external validity in predicting mortality from traumatic brain injury. \emph{Journal of clinical epidemiology}. 2016;78:83-89.}

\bibitem[\citeproctext]{ref-gravesteijn2020machine}
\CSLLeftMargin{25. }%
\CSLRightInline{{Gravesteijn BY, Nieboer D, Ercole A, et al.} Machine learning algorithms performed no better than regression models for prognostication in traumatic brain injury. \emph{Journal of clinical epidemiology}. 2020;122:95-107.}

\bibitem[\citeproctext]{ref-steyerberg2003internal}
\CSLLeftMargin{26. }%
\CSLRightInline{Steyerberg EW, Bleeker SE, Moll HA, Grobbee DE, Moons KG. Internal and external validation of predictive models: A simulation study of bias and precision in small samples. \emph{Journal of clinical epidemiology}. 2003;56(5):441-447.}

\bibitem[\citeproctext]{ref-steinley2003local}
\CSLLeftMargin{27. }%
\CSLRightInline{Steinley D. Local optima in k-means clustering: What you don't know may hurt you. \emph{Psychological methods}. 2003;8(3):294.}

\bibitem[\citeproctext]{ref-von2010clustering}
\CSLLeftMargin{28. }%
\CSLRightInline{{Von Luxburg U et al.} Clustering stability: An overview. \emph{Foundations and Trends{®} in Machine Learning}. 2010;2(3):235-274.}

\bibitem[\citeproctext]{ref-maier2013result}
\CSLLeftMargin{29. }%
\CSLRightInline{Maier M, Von Luxburg U, Hein M. How the result of graph clustering methods depends on the construction of the graph. \emph{ESAIM: Probability and Statistics}. 2013;17:370-418.}

\end{CSLReferences}
\end{document}